\documentclass[a4paper,fleqn,usenatbib]{mnras}

\usepackage[T1]{fontenc}
\usepackage{ae,aecompl}
\usepackage[normalem]{ulem}

\usepackage{graphicx}	
\usepackage{amsmath}	
\usepackage{amssymb}
\usepackage{subcaption}
\usepackage{adjustbox}

\title [Characteristic emission of the blazar Mkn\,421]{Do radiative losses determine the characteristic emission of the blazar Mkn\,421?}

\author[C Baheeja et al.]{
C. Baheeja$^{1}$\thanks{E-mail: baheeja314@gmail.com}, S. Sahayanathan$^{2,3}$\thanks{E-mail: sunder@barc.gov.in}, Frank M. Rieger$^{4,5}$, Sitha K Jagan  $^{1}$  and \newauthor C. D. Ravikumar$^{1}$
\\
\textit{$^{1}$ Department of Physics, University of Calicut, Malappuram, Kerala, India}\\
\textit{$^{2}$ Astrophysical Sciences Division, Bhabha Atomic Research Centre, Mumbai - 400085, India}\\
\textit{$^{3}$ Homi Bhabha National Institute, Mumbai 400094, India}\\
\textit{$^{4}$ Institut f$\ddot{u}$r Theoretische Physik (ITP), Universit$\ddot{a}$t Heidelberg, Philosophenweg 12, 69120 Heidelberg, Germany}\\
\textit{$^{5}$ Max-Planck-Institut f$\ddot{u}$r Kernphysik, P.O. Box 103980, 69029 Heidelberg, Germany}
}

\date{Accepted XXX. Received YYY; in original form ZZZ}

\pubyear{2022}

\begin{document}
\label{firstpage}
\pagerange{\pageref{firstpage}--\pageref{lastpage}}
\maketitle

\begin{abstract}
	The radiative loss interpretation for the broken power-law spectra of blazars is often questioned since the difference 
	between the indices does not support this inference. Using the blazar Mkn\,421 as a case study, we performed a detailed 
	analysis of its characteristic photon energy where the spectral index changes significantly. We used the observations 
	of the source by \emph{Swift}-XRT from 2008 to 2019 to identify the characteristic photon energy and the corresponding 
	spectral indices. The spectra in the energy range 0.3--10.0 keV can be well fitted by a log parabola as well as a 
	smooth broken power-law. 
          From the smooth broken power-law spectral fit we show that the spectral indices before and after the characteristic 
          photon energy are strongly anti-correlated. Further, the spectral curvature 
	measured at the characteristic photon energy indicates an anti-correlation with the low energy spectral index while 
	the high energy spectral index shows a positive correlation.
	These findings are at variance with a simple radiative loss interpretation for the characteristic photon energy, and 
	alternative scenarios are thus discussed. Though these scenarios are in principle capable of reproducing the 
	correlation results, they deviate significantly from the observed properties.	
\end{abstract}




\section{Introduction}

The BL Lac object Mkn\,421 is one of the nearest ($z=0.031$) and extensively studied blazars with a 
non-thermal spectrum extending from radio to very high energy gamma rays \citep{2009ApJ...703..169A,
2011ApJ...736..131A,2012A&A...542A.100A,2016ApJ...819..156B,2016ApJS..222....6B}. The source 
exhibits rapid flux and spectral variability, suggesting the emission to originate from a relativistic moving jet 
aligned close to the line of sight \citep{1995MNRAS.273..583D}. Its broadband spectral energy distribution 
(SED) is characterized by two broad peaks, where the low-energy component is interpreted as synchrotron emission 
from a relativistic electron population \citep{10.1111/j.1365-2966.2007.12758.x} while the high-energy component 
is usually attributed to inverse Compton scattering of the synchrotron photons by the same electron population 
\citep[e.g.,][]{2012A&A...542A.100A,2011ApJ...736..131A,2011ApJ...738...25A}. This synchrotron self Compton 
(SSC) interpretation of the high energy emission is further strengthened by the correlated X-ray -- TeV flux variability 
with quadratic dependence \citep[e.g.,][]{1997A&A...320...19M,2005ApJ...630..130B,2007A&A...462...29G,
2008ApJ...677..906F,2010A&A...510A..63K,2011ApJ...734..110B,2018ApJ...854...66K,2018ApJ...858...68K}.

In general, different types of non-thermal electron distributions, mostly involving a power-law regime 
(e.g., broken power-law, power-law with exponential cut-off etc.), have been used to account 
for the broadband SED of blazars \citep[e.g.,][]{2017ApJ...836...83S,2016ApJ...827...55K,Yan_2013,
2007A&A...467..501T,Krawczynski_2004}.

On the other hand, since the narrow band X-ray spectrum can deviate significantly from a simple power-law, the photon number distribution is often represented by a log-parabolic function 
\begin{align}\label{eq:logpar}
	N(\epsilon)\propto \left(\frac{\epsilon}{\epsilon_0}\right)^{-\alpha-\beta\,\rm{log}(\epsilon/\epsilon_0)}\quad {\rm .}
\end{align}
Here, $\epsilon$ is the photon energy, $\alpha$ is the spectral slope at energy $\epsilon_0$ and $\beta$ 
is the spectral curvature defined at the peak energy $\epsilon_p$ of the SED \citep{2004A&A...413..489M,
2007A&A...467..501T}. 
In case of Mkn\,421, $\epsilon_p$ falls into the soft X-ray regime and the hard X-ray energy band 
probes the decline of the synchrotron spectral component. The spectrum at these energy regimes 
individually can be reproduced by a log-parabolic function, though this approach is not very successful in 
explaining the combined optical/UV and X-ray spectra \citep{2004A&A...413..489M,2009A&A...501..879T,
2015A&A...580A.100S}

The characteristic peak photon energy ($\epsilon_p$) of Mkn\,421, at which the synchrotron spectral 
component carries maximum power, can vary significantly depending on the source state. For instance, 
\cite{Kapanadze_2020} showed that $\epsilon_p$ varies from  $< 0.1$ to  $>15$ keV during different
flux states using Swift-XRT observations. Such large variation in $\epsilon_p$ is also reported by various 
authors using \emph{RXTE}, \emph{BeppoSAX}, \emph{XMM-Newton} observations 
\citep[e.g.,][]{2004A&A...413..489M,2007A&A...466..521T}. The spectral shape around $\epsilon_p$ is 
usually well represented by a log-parabola function. Correlation studies between the fit parameters on
the other hand, are often contradictory or inconclusive.  For example, $\epsilon_p$ and $\beta$ obtained 
through the log-parabolic spectral fit of \emph{BeppoSAX}, \emph{RXTE}, \emph{XMM-Newton} and 
\emph{Swift}-XRT observations of Mkn\,421 during 1997--2006 were found to be anti-correlated 
\citep{2007A&A...466..521T,2008A&A...478..395M,2009A&A...501..879T,2011ApJ...739...66T}. 
However, no significant correlation between these quantities was witnessed in case of \emph{Swift}-XRT 
observation during 2005--2008 \citep{2018ApJ...854...66K}, 2009--2012 \citep{2018ApJ...858...68K}, 
2015--2018 \citep{Kapanadze_2020}, and also in combination with \emph{Nu}STAR observations during 
April 2013 \citep{Sinha_2015}. 
Similarly, \cite{2004A&A...413..489M} have reported a strong positive correlation between the spectral 
index at 1 keV and the curvature parameter using \emph{BeppoSAX} observations (0.1$-$100 keV) 
of Mkn\,421 during May 1999 and April$-$May 2000; however, \emph{Swift}-XRT observations and 
combined \emph{Swift}-XRT/\emph{Nu}STAR spectra did not show an appreciable correlation \citep{Sinha_2015,
Kapanadze_2017, 2018ApJ...854...66K}.
In spite of this seemingly contradictory results, in many observations the enhancement in flux is associated 
with a spectral hardening commonly referred to as ``harder when brighter'' trend \citep{2016ApJ...831..102K,
Kapanadze_2017, 2018ApJ...854...66K,2018ApJ...858...68K,Kapanadze_2020}.

The power-law/log-parabolic photon spectra demand the emitting electron distribution to be also 
power-law/log-parabola type \citep{1962SvA.....6..317K,2004A&A...413..489M,2009A&A...501..879T}. 
A power-law electron distribution can be readily achieved under Fermi acceleration
\citep[e.g.,][]{2007Ap&SS.309..119R}, while a log-parabolic electron distribution may indicate  
an energy-dependence in the particle acceleration and/or diffusion process \citep{2004A&A...413..489M,
2018MNRAS.478L.105J,10.1093/mnras/sty2003,2020MNRAS.499.2094G}. In particular, when the 
electron escape time-scale from the main particle acceleration site becomes mildly energy-dependent (referred
to as `energy dependent diffusion model', hereafter EDD), the resultant electron distribution has been shown 
to follow a log-parabola function \citep{2018MNRAS.478L.105J}. 
On the other hand, when the energy-dependence is strong, the shape deviates from a log-parabola. 
The latter case is witnessed in the hard X-ray spectra of Mkn\,421 
\citep{10.1093/mnras/sty2003,2020MNRAS.499.2094G} observed by \emph{Nu}STAR. The spectra 
corresponding to different flux states can be well fitted by synchrotron emission originating from an 
electron distribution obtained in an EDD model with a strong energy-dependent escape time scale. 
Incidentally, the hard X-ray spectra can also be explained by a log-parabolic electron distribution and 
hence, single model fits of the source spectra does not allow one to differentiate between these two models. 
However, the fit parameters of the EDD model corresponding to the different flux states indicate a 
strong correlation, while this is not the case with a log-parabolic electron distribution.

The peak spectral energy $\epsilon_p$ can be translated to a break or peak energy, $E_b= 
\gamma_b m_e c^2$, in the emitting electron energy distribution\footnote{The electron energy 
distribution is represented by $E^2 N(E)$, where $N(E)$ is the specific electron number density. 
Most of the electron energy resides at $E_b$ when the broken power law indices are $>-2$ and 
$<-2$, respectively.} under synchrotron theory. The system deposits most of its power to the electrons 
at this energy, and hence identifying the process that determines  $\gamma_b$ (or $\epsilon_p$) is 
important to understand the dynamics of the source. Radiative cooling (synchrotron and/or inverse 
Compton) of a power-law electron distribution can give rise to a broken power law electron distribution 
with power-law indices differing by unity \citep{1962SvA.....6..317K}. Correspondingly, the photon 
spectrum will also be a broken power-law with a difference of $0.5$ between high and low energy spectral 
indices \citep{1986rpa..book.....R}.
However, SED modelling of blazars often does not support this inference and the index difference is frequently 
found to be significantly larger than $0.5$ \citep[e.g.,][]{Mankuzhiyil_2012}. On the other hand, radiative 
cooling of a log-parabolic electron distribution can transform the distribution into a broken log-parabolic form. 
This could probably explain the large differences between spectral indices \citep{2018MNRAS.478L.105J}. 
Under this interpretation, the increase or decrease in the spectral slope at low energies ($<\epsilon_p$) will 
be associated with a similar spectral slope change at high energies ($>\epsilon_p$). 
In principle, the assumption of multiple acceleration processes is also capable of producing broken power-law 
electron distributions \citep[e.g.,][]{2008MNRAS.388L..49S}. The low and high energy indices would then be 
governed by the associated acceleration rates.

In the present work, we examine the radiative loss origin of $\epsilon_p$ in the high energy peaked blazar 
Mkn\,421. For this, we use long term observations of the source by \emph{Swift}-XRT spanning from January 
2008 to December 2019. The spectral resolution of \emph{Swift}-XRT is appreciable within the energy range 
$0.3$ to $10.0$ keV, encompassing a broad range of $\epsilon_p$, and hence, represents an suitable experiment 
to perform this study. The X-ray spectrum is well represented by a log-parabola function which allows one to
constrain $\epsilon_p$. The spectral indices before and after $\epsilon_p$ are obtained by refitting the spectrum 
with a smooth broken power-law. We then study the correlation between these quantities to explore the possible origin 
of $\epsilon_p$.

The paper is organised as follows: In Section 2, we discuss the observation and data reduction procedure, 
while the X-ray spectral study and correlations between fitting parameters are described in Section 3. Discussion 
and summary of the work are presented in Section 4.

\section{Observation and Data Analysis}
We analysed the \emph{Swift}-XRT \citep{2005SSRv..120..165B} observations of Mkn 421 during 2008--2019. 
The data were retrieved from NASA's HEASARC interface and processed using XRTDAS software package 
(Version 3.0.0) included within HEASOFT package (Version 6.22.1). We used only those observations which 
were performed in Windowed Timing (WT) mode and selected events with 0--2 grades. Standard procedures 
employing \emph{xrtpipeline} (Version v0.13.4) were used for calibration and data cleaning.

A circular region of about  20--30 pixel radius centred at the source was used to extract the source spectrum, while 
about 30--40 pixel radius circular region which is free from the source contamination was used to extract the background 
spectrum. For observations with pileup, the source spectra were extracted by excluding the central circular aperture of 
2--3 pixel radius. The final spectrum was produced using \emph{xrtproducts} (Version v0.4.2). The ancillary response 
files (ARFs) were generated using the XRTMKARF task and the latest response matrices files (RMF) were used from 
the \emph{Swift} CALDB. To ensure better $\chi^2$ statistics, the spectra were grouped to 20 photons per bin using 
the tool GRPPHA v.3.0.1. We rejected some observations that were strongly biased due to the dead columns on the CCD.

\section{X--ray spectral fit}
The X-ray spectra in the  $0.3-10.0$ keV energy band encompass or are close to $\epsilon_p$, and hence are 
significantly curved. The reduced data is analysed with the XSPEC package \citep{1996ASPC..101...17A} using a 
log-parabolic function while fixing the hydrogen column density to the Galactic value $N_H = 1.92\times 10^{20} \rm
cm^{-2}$ \citep{2005A&A...440..775K} for all the observations. Since our aim is to understand the origin of 
$\epsilon_p$, we used the \emph{eplogpar} model \citep{2009A&A...501..879T} to perform the spectral fit. The 
log-parabolic function under this model is expressed in terms of $\epsilon_p$ as

\begin{align}
	N(\epsilon) \propto 10^{-\beta\,\left({\rm log}(\epsilon/\epsilon_p)\right)^2}/\epsilon^2.
\end{align}

\begin{figure}
\includegraphics[width=1.0\linewidth ]{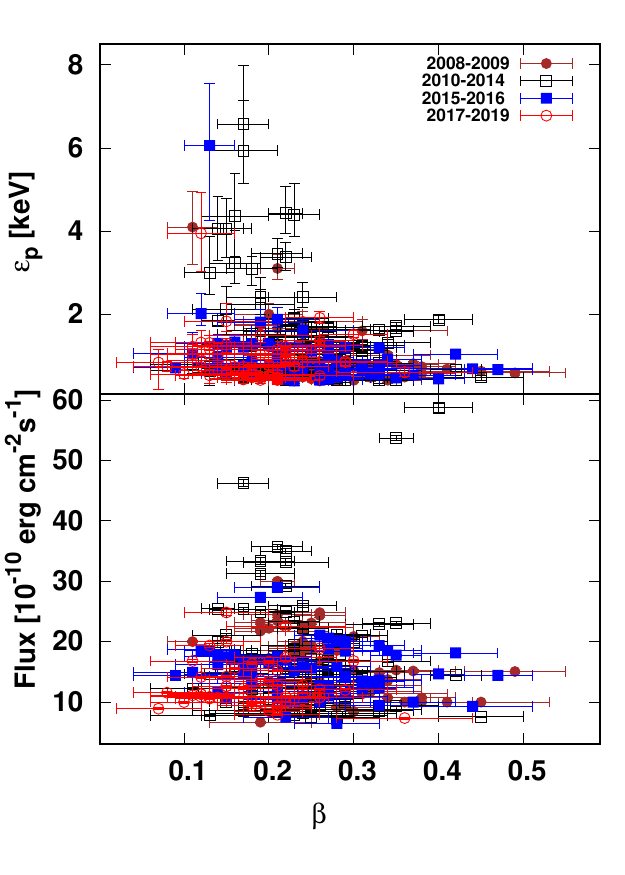}
\caption{The scatter plot between $\beta$ and $\epsilon_p$ obtained from \emph{eplogpar} (top panel) and between $\beta$ 
and the integrated 0.3--10.0 keV flux (lower panel) for different epochs.}
\label{fig:ep-beta}
\end{figure}

We excluded observations having large reduced chi-squared ($\chi_{\rm red}$>1.2); however, for some 
observations we have extracted spectra from individual orbits to obtain a better fit. 
Certain observations corresponding to a single orbit were also divided segment-wise to improve the spectral fit. 
The details of the observation are shown in Table \ref{tab:spectral-fitting}. For many observations, the obtained 
$\epsilon_p$ lies outside the $0.3-10.0$ keV range. These estimates are not reliable, since they fall outside the 
spectral energy range of \emph{Swift}-XRT  and are thus excluded. This leaves 258 spectra with 
$\epsilon_p$ between $0.4$ to $7.0$ keV for the present study. 
Consistent with the literature \citep{2004A&A...413..489M,2007A&A...466..521T,2016ApJ...831..102K, 
Kapanadze_2017}, the spectra in this energy regime can be well fitted by a log-parabola. In Table \ref{tab:spectral-fitting} 
(column 4$-$5), we provide the fit results and in Figure \ref{fig:ep-beta} (upper panel) we 
show the scatter plot between $\epsilon_p$ and $\beta$. The Spearman rank correlation study between $\epsilon_p$ and 
$\beta$ yields $r=-0.28$ with a null hypothesis probability of $1.35 \times 10^{-05}$. This result is consistent with earlier 
studies where a weak or no correlation was observed \citep{2018ApJ...854...66K,2018ApJ...858...68K, Sinha_2015}. 

\begin{figure}
\includegraphics[width=1.2\linewidth]{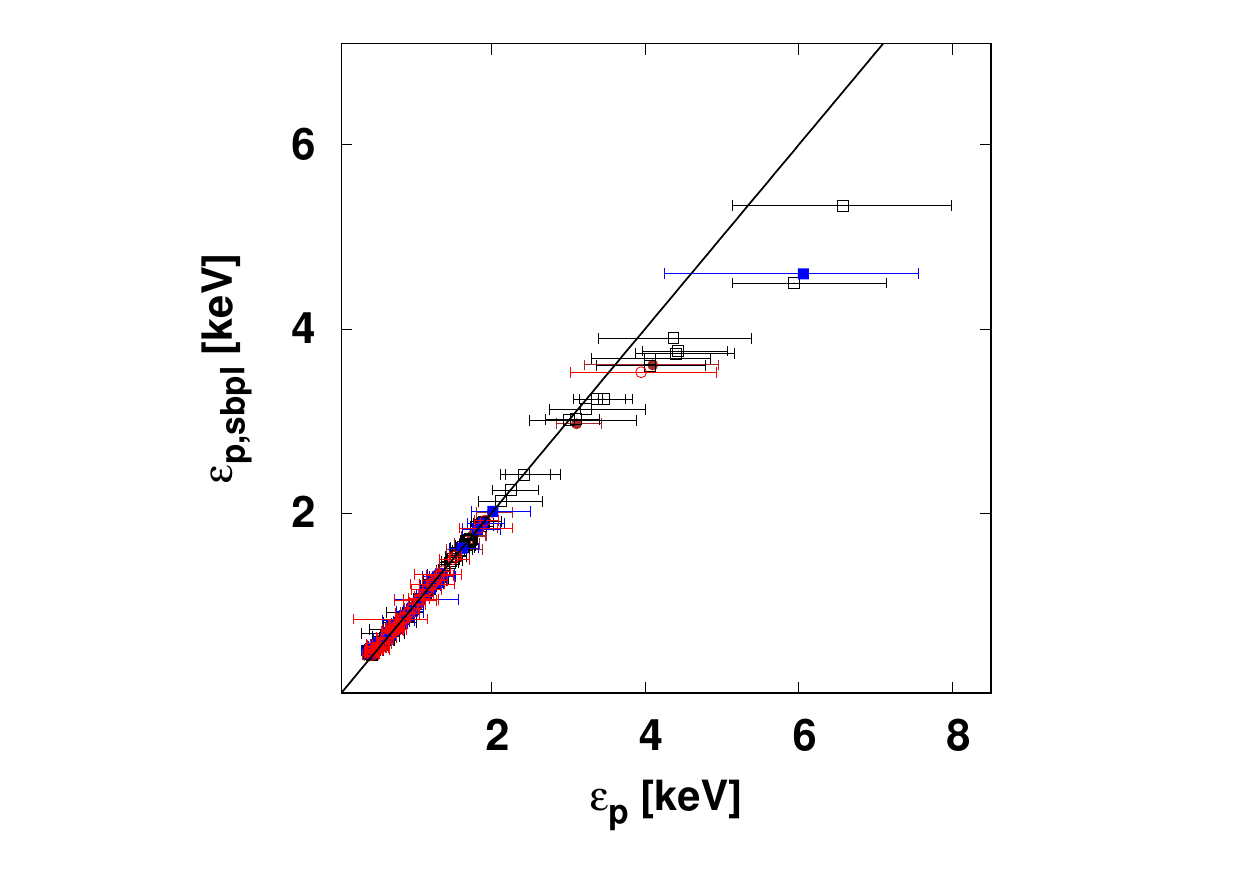}
\caption {The scatter plot between $\epsilon_{p}$ and $\epsilon_{p,sbpl}$ along with the identity line (see text). The symbols are as shown in Figure \ref{fig:ep-beta}.}
\label{fig:logpar-ep-sbpl-ep-xrt}
\end{figure}

\begin{figure}
\includegraphics[width=1.2\linewidth]{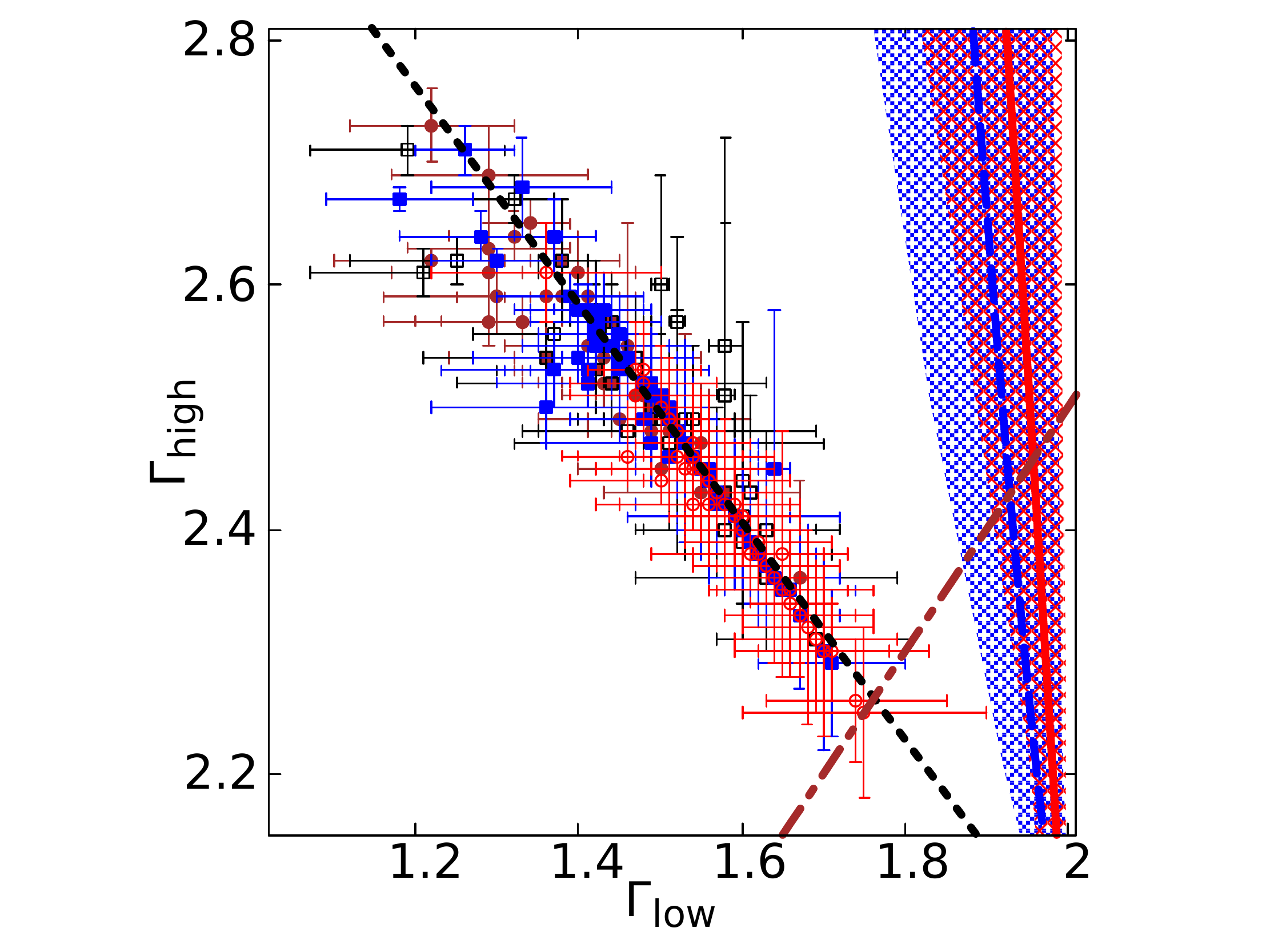}
\caption {Scatter plot beween $\Gamma_{\rm low}$ and $\Gamma_{\rm high}$: black dotted line represents the best fit line of 
the observational $\Gamma_{\rm low}$ and  $\Gamma_{\rm high}$. The red solid line correspond to the synchrotron spectrum of an electron distribution of power-law with an exponential cut-off for 1.75 $<\psi<$ 2.0, and the red shaded region represents its 1-$\sigma$ deviation. The blue dashed line correspond to the synchrotron spectrum of a electron distribution with a gradual decline at the maximum 
electron energy by varying $\phi$ between 1.3 and 2.0, and the blue shaded region represents its 1-$\sigma$ deviation. The brown dotted dash line represents the cooling break. The symbols are as shown in Figure \ref{fig:ep-beta}.}
\label{fig:theoretical-fit}
\end{figure}

\begin{figure}
\includegraphics[width=1.4\linewidth]{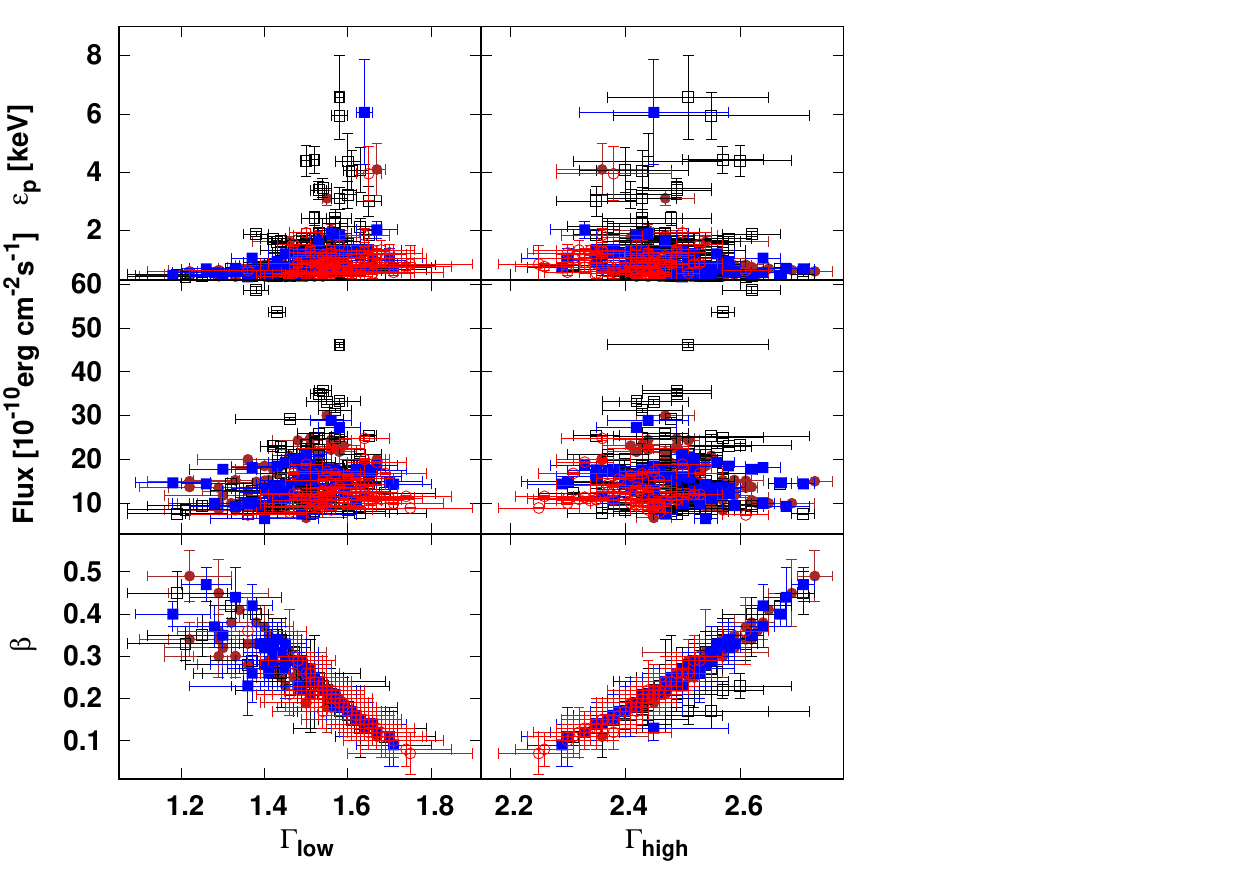}
\caption {Left panel: scatter plot between $\Gamma_{\rm low}$ and curvature parameter $\beta$, 0.3--10.0 keV flux and $\epsilon_p$; Right panel: 
scatter plot between $\Gamma_{\rm high}$ and curvature parameter $\beta$, 0.3--10.0 keV flux and $\epsilon_p$. The symbols are as shown in Figure \ref{fig:ep-beta}. }
\label{fig:mutiplot-3}
\end{figure}

The radiative loss interpretation of $\epsilon_p$ predicts the spectral slopes at energies $\epsilon \ll \epsilon_p$ 
and $\epsilon \gg \epsilon_p$ to be positively correlated. To examine this, we fitted the spectra with a smooth broken 
power-law function defined by

\begin{align}\label{eq:sbpl}
	N(\epsilon)\propto \left[ \left(\frac{\epsilon}{\epsilon_{\rm b}}\right)^{\Gamma_{\rm low}}+\left(\frac{\epsilon}{\epsilon_{\rm b}}\right)^{\Gamma_{\rm high}}\right]^{-1} 
\end{align}
where $\Gamma_{\rm low}$ and $\Gamma_{\rm high}$ are the indices before and after the break energy $\epsilon_b$. 
The peak of the smooth broken power-law function in $\epsilon^2\, N(\epsilon)$ representation is
\begin{align}\label{eq:sbpl-ep}
	\epsilon_{\rm p,sbpl} = \epsilon_{\rm b} \left(\frac{\Gamma_{\rm high}-2}{2-\Gamma_{\rm low}}\right)^{\frac{1}{\Gamma_{\rm low}-\Gamma_{\rm high}}}\,.    
\end{align}
For typical values of $\Gamma_{\rm high}$ ($\sim$ 2.5) and $\Gamma_{\rm low}$ ($\sim$ 1.5), $\epsilon_{\rm b}\approx \epsilon_{\rm p,sbpl}$.
The function (\ref{eq:sbpl}) was added as a local model (\emph{sbpl}) in XSPEC and the $0.3-10.0$ keV 
\emph{Swift}-XRT observations of Mkn\,421 were refitted.  
However, the narrow-band X-ray spectra do not allow us to sufficiently constrain all parameters of the model.
Hence, we performed a fitting with $\epsilon_b$ fixed to the value $\epsilon_p$ obtained from the \emph{eplogpar} 
spectral fit. A subsequent inspection shows that $\epsilon_{p,sbpl}$ estimated from equation (\ref{eq:sbpl-ep}) 
using the so obtained best fit $\Gamma_{\rm low}$ and $\Gamma_{\rm high}$ does not differ much from 
$\epsilon_{p}$ (Figure \ref{fig:logpar-ep-sbpl-ep-xrt}), suggesting that our choice of $\epsilon_b$ does not 
strongly affect the outcome.
We consider only those \emph{Swift}-XRT observations with $0.4$ keV $<\epsilon_p<$ $7.0$ keV since estimation of 
$\Gamma_{\rm low}$ and $\Gamma_{\rm high}$ demands $\epsilon_p$ to be within the spectral energy range of 
\emph{Swift}-XRT.
The resultant best fit values for $\Gamma_{\rm low}$ and $\Gamma_{\rm high}$ are given in Table \ref{tab:spectral-fitting} 
(column 7$-$8), while the scatter plot between these quantities is shown in Figure \ref{fig:theoretical-fit}.
The correlation study between these quantities yields a linear correlation coefficient $r = -0.95$ with significance, $p > 99.99$ 
per cent \citep{1992nrfa.book.....P}. This strong anti-correlation between $\Gamma_{\rm low}$ and $\Gamma_{\rm high}$ 
poses a serious challenge to a simple radiative cooling interpretation of $\epsilon_p$. A linear fit to $\Gamma_{\rm low}$ 
and $\Gamma_{\rm high}$  results in 
\begin{align}\label{eq:fitline}
 \Gamma_{\rm high} = (-0.89 \pm 0.05) \Gamma_{\rm low} + (3.83 \pm 0.08)
\end{align} with a goodness of fit (q-value) of $0.99$.
The inability of the simple radiative loss interpretation to successfully account for the origin of $\epsilon_p$ can 
be further visualized by studying the correlation between $\Gamma_{\rm low}$ or $\Gamma_{\rm high}$ with the 
curvature parameter $\beta$ obtained from \emph{eplogpar} model. In Figure \ref{fig:mutiplot-3} (bottom panel), 
we show the scatter plot between these quantities.  The strong linear correlation of $\Gamma_{\rm high}$ with 
$\beta$ (r = 0.96, p > 99 per cent) and anti-correlation of $\Gamma_{\rm low}$ with $\beta$ (r = -0.92, p > 99 per cent) 
are  consistent with the anti-correlation observed between $\Gamma_{\rm low}$ and $\Gamma_{\rm high}$. A softening of 
the high energy index is associated with a hardening of the low 
energy index (broadening of the synchrotron spectral component). Correspondingly, high curvature $\beta$ will 
be associated with hard $\Gamma_{\rm low}$ and steep $\Gamma_{\rm high}$ (narrowing of the synchrotron 
spectral component).

We did not find any appreciable correlation between $\Gamma_{\rm low}$ or $\Gamma_{\rm high}$ with the
$0.3-10.0$ keV integrated flux ($F_{0.3-10.0\rm keV}$, middle panel of Figure \ref{fig:mutiplot-3}).
This indicates that the narrowing of the synchrotron component of the SED of Mkn\,421 is not associated with 
flux state. This is also consistent with the weak correlation observed between $\beta$ and $F_{0.3-10.0 \rm keV}$ 
(lower panel of Figure \ref{fig:ep-beta}). Similarly, no significant correlation is witnessed between $\epsilon_p$ and 
$\Gamma_{\rm low}$ or $\Gamma_{\rm high}$ (top panel of Figure \ref{fig:mutiplot-3}). Hence, the narrowing 
of the SED cannot be attributed to ``bluer when brighter'' trend of Mkn\,421.

\begin{table*}
\begin{center}
\caption{Best fit parameters of spectral fitting using \emph{eplogpar} model and \emph{sbpl} model} 
\label{tab:spectral-fitting}         
\begin{tabular}{l c c c c c c c c c c}
\hline
 ObsID	&	Date of Obs.	&	Exposure	&		&	\emph{eplogpar}	&		&		&	\emph{sbpl}	&	 & 	Flux$_{0.3-10.0\,\rm kev}$\\
 
 \cline{4-6}
 \cline{7-9}
 
  	&		&	(sec)	&	$\epsilon_p$~(keV)	&	$\beta$	&	
  	$\chi_{red}^2$ (dof) 	&	$\Gamma_{\rm low}$	&	$\Gamma_{\rm high}$		&	$\chi_{red}^2$ (dof)  & ($ 10^{-10} \rm erg\,cm^{-2} s^{-1}$)	\\
  	(1)	&	(2)	&	(3)	&	(4)	&	(5)	&	(6)	&	(7)		&	(8) & (9)&(10)	\\

\hline
\hline

30352053-Orb2	&	2008-01-16	&	575.1	&	0.42$^{+0.07}_{-0.08}$&	0.28$\pm$0.05	&	1.04 (305)	&	1.36$\pm$0.12	&	2.54$\pm$0.02	&	1.06 (305)	&	20$\pm$0.28	\\
30352054	&	2008-01-16	&	1134.085	&	0.42$^{+0.04}_{-0.05}$	&	0.3$\pm$0.03	&	1.08 (394)	&	1.33$\pm$0.1	&	2.57$\pm$0.01	&	1.12 (394)	&	15.26$\pm$0.15	\\
30352056	&	2008-01-17	&	944.126	&	0.43$^{+0.05}_{-0.07}$	&	0.34$\pm$0.04	&	0.95 (322)	&	1.29$\pm$0.12	&	2.61$\pm$0.02	&	1.0 (322)	&	13.54$\pm$0.17	\\
30352055-Orb1	&	2008-01-17	&	394.9	&	0.64$^{+0.07}_{-0.08}$	&	0.45$\pm$0.08	&	1.11 (227)	&	1.29$\pm$0.12	&	2.69$\pm$0.04	&	1.12 (227)	&	10$\pm$0.21	\\
30352055-Orb2	&	2008-01-17	&	362.6	&	0.60$^{+0.05}_{-0.04}$	&	0.49$\pm$0.06	&	1.09 (260)	&	1.22$\pm$0.1	&	2.73$\pm$0.03	&	1.12 (260)	&	15.04$\pm$0.26	\\
30352058	&	2008-01-18	&	889.108	&	0.57$^{+0.04}_{-0.04}$	&	0.33$\pm$0.04	&	1.07 (378)	&	1.38$\pm$0.07	&	2.59$\pm$0.02	&	1.1 (378)	&	14.88$\pm$0.16	\\
30352059	&	2008-01-19	&	919.112	&	0.43$^{+0.04}_{-0.04}$	&	0.34$\pm$0.04	&	1.04 (359)	&	1.22$\pm$0.12	&	2.62$\pm$0.01	&	1.06 (359)	&	13.73$\pm$0.15	\\
30352060	&	2008-02-06	&	753.06	&	0.60$^{+0.05}_{-0.06}$	&	0.26$\pm$0.03	&	1.13 (429)	&	1.48$\pm$0.06	&	2.51$\pm$0.02	&	1.15 (429)	&	24.33$\pm$0.24	\\
30352066	&	2008-02-10	&	1174.1	&	0.74$^{+0.03}_{-0.03}$	&	0.3$\pm$0.03	&	1.17 (467)	&	1.46$\pm$0.04	&	2.55$\pm$0.02	&	1.18 (467)	&	20.85$\pm$0.18	\\
30352068	&	2008-02-11	&	1868.208	&	0.46$^{+0.03}_{-0.05}$	&	0.25$\pm$0.02	&	1.1 (495)	&	1.44$\pm$0.06	&	2.52$\pm$0.01	&	1.14 (495)	&	18.61$\pm$0.14	\\
\hline
\end{tabular}
\end{center}
\begin{flushleft}
	\vspace{-2mm}
	\footnotesize
(Table is available in its entirety in the online version.)
\end{flushleft}
\end{table*}

\section{Discussion and Summary}

The observed anti-correlation between $\Gamma_{\rm low}$ and $\Gamma_{\rm high}$ is inconsistent with a 
simple radiative loss interpretation of $\epsilon_p$. 
To examine alternate explanations, we consider the case 
where the synchrotron spectral component is governed by a power-law with an exponential cut-off. Such spectral 
shape could be possible when the underlying particle distribution has a sharp cut-off at the maximum achievable
electron energy. The high-energy end of the synchrotron spectrum will then be governed by the exponential 
decay of the synchrotron single particle emissivity function. To mimic this, we assume a synchrotron spectral 
shape as 
\begin{align}
	N(\epsilon) \propto \epsilon^{-\psi}{\rm exp}\left(-\frac{\epsilon}{\epsilon_{c}}\right) \,.
\end{align}
The SED peak in $\epsilon^2\, N(\epsilon)$ representation will be
\begin{align}
	\epsilon_{p,\rm exp} = (2-\psi)\epsilon_{c}\,,
\end{align}  
so that the spectral slope of the photon distribution around $\epsilon_{p,\rm exp}$ 
can be expressed as

\begin{align}\label{eq:gamma_exp}
	\Gamma_{\rm exp} (\epsilon)= \psi+\frac{\epsilon}{\epsilon_{p,\rm exp}}\left(2-\psi\right)\,.
\end{align}

In this case, the spectral slopes at energies $\epsilon<\epsilon_{p,\rm exp}$ and $\epsilon >\epsilon_{p,\rm exp}$ 
will be anti-correlated.  If we set $\epsilon_{p,\rm exp}  = 1.14$ keV corresponding to the average value 
of $\epsilon_p$ estimated from all the observations, the dependence of 
 $\Gamma_{\rm exp}(0.3 \,{\rm keV}) = \Gamma_{\rm low}$ and $\Gamma_{\rm exp}(10.0 \,{\rm keV})
=\Gamma_{\rm high}$ can be studied for different values 
of $\psi$ (1.75 $<\psi<$ 2.0). In Figure \ref{fig:theoretical-fit}, we show this dependence as red solid 
line and the red shaded region represents its 1-$\sigma$ deviation. Though this interpretation supports the 
anti-correlation between the indices, it deviates largely from the best fit 
line obtained from the observed $\Gamma_{\rm low}$ and $\Gamma_{\rm high}$ (using \emph{sbpl}). 
\emph{Nu}STAR observations of the source also do not support such an interpretation \citep{2016ApJ...819..156B}.

Another possible explanation for $\epsilon_p$ could be, when the underlying particle distribution exhibits 
a gradual decline at the maximum electron energy rather than a sharp cut-off, that dominates the spectral 
shape around $\epsilon_p$. To examine this possibility, we use the electron distribution accelerated at a 
shock front with constant acceleration and escape time scales \citep{1998A&A...333..452K}
\begin{align}\label{eq:partcurv}
N(\gamma) \propto \gamma^{-\left(1+\phi\right)} \left(1-\frac{\gamma}{\gamma_{\rm max}}\right)^{\phi-1}\,.
\end{align}
Here, $\gamma$ is the Lorentz factor of the electron, $\gamma_{\rm max}$ is the Lorentz factor corresponding 
to the maximum electron energy, and $\phi$ is the ratio of the acceleration and escape time scales.
The observed synchrotron spectrum will then be \citep{1986rpa..book.....R}
\begin{align}\label{eq:syn_emiss}
	N(\epsilon)\propto \frac{1}{4\pi}\int\limits_{\gamma_{\rm min}}^{\gamma_{\rm max}} 
	P_{\rm syn}(\gamma,\epsilon^*)\,N(\gamma)\,d\gamma
\end{align}
where $P_{\rm syn}$ is the synchrotron single particle emissivity and $\epsilon^*$ is the photon energy in the
comoving frame of the emitting region carried by the blazar jet \citep[e.g.,][]{RevModPhys.56.255}. The 
integration in equation (\ref{eq:syn_emiss}) is performed numerically using quadrature, and we study the spectral 
shape around the peak. Again setting $\epsilon_p$ at 1.14 keV, 
the spectral index at 0.3 keV and 10 keV can be studied for different values of $\phi$ (1.3 $<\phi<$ 2.0). 
In Figure \ref{fig:theoretical-fit}, we show the plot between these indices as blue solid line and the blue shaded 
region represents the 1-$\sigma$ deviation of the line. This interpretation for $\epsilon_p$ also supports an 
anti-correlation between $\Gamma_{\rm low}$ and $\Gamma_{\rm high}$; however, it still deviates significantly 
from the best fit line.

An alternative scenario which could be capable of reproducing the best fit straight line besides explaining the 
observed anti-correlation between $\Gamma_{\rm low}$ and $\Gamma_{\rm high}$ is when the synchrotron 
spectral component is a superposition of multiple broken power-law components. This would implicitly assume 
that the emission region is not reducible to a single homogeneous zone. The dominant broken power-law 
component then determines $\epsilon_p$ and with proper choice of break energy and/or the normalisation 
one could possibly reproduce the observed best fit straight line. However, unless some fine tuning occurs, one 
would also expect $\epsilon_p$ or the flux to be correlated or anti-correlated with $\Gamma_{\rm low}$ or 
$\Gamma_{\rm high}$. The apparent absence of such correlations (top and middle panels of 
Figure \ref{fig:mutiplot-3}) would again seem to disfavour simple variants of such a scenario.

As shown above, the curvature in the electron distribution introduced by the simplistic assumptions of constant 
acceleration and escape time-scales is unable to account for the observed properties. Probably, relaxing these 
assumptions, by including more complex energy-dependent acceleration and escape time-scales, could modify 
the correlation and reduce the deviation.

Multiple particle acceleration scenarios are also capable, in principle, of producing electron distributions that 
could imitate a broken power-law \citep[e.g.,][]{2008MNRAS.388L..49S}. 
In this case, the indices are governed by the ratio of the acceleration and escape timescales of the associated 
acceleration processes.  The dominant acceleration processes in the blazar jet are assumed to be shock- and 
stochastic-type. Since the rate of acceleration by both these processes depends on the nature of particle 
diffusion into the jet medium, the corresponding particle power-law indices may be correlated. However, the 
exact nature of this correlation and comparison with observations would demand a detailed study/simulation of
AGN jets considering these acceleration processes.

\begin{figure}
\includegraphics[width=1.2\linewidth]{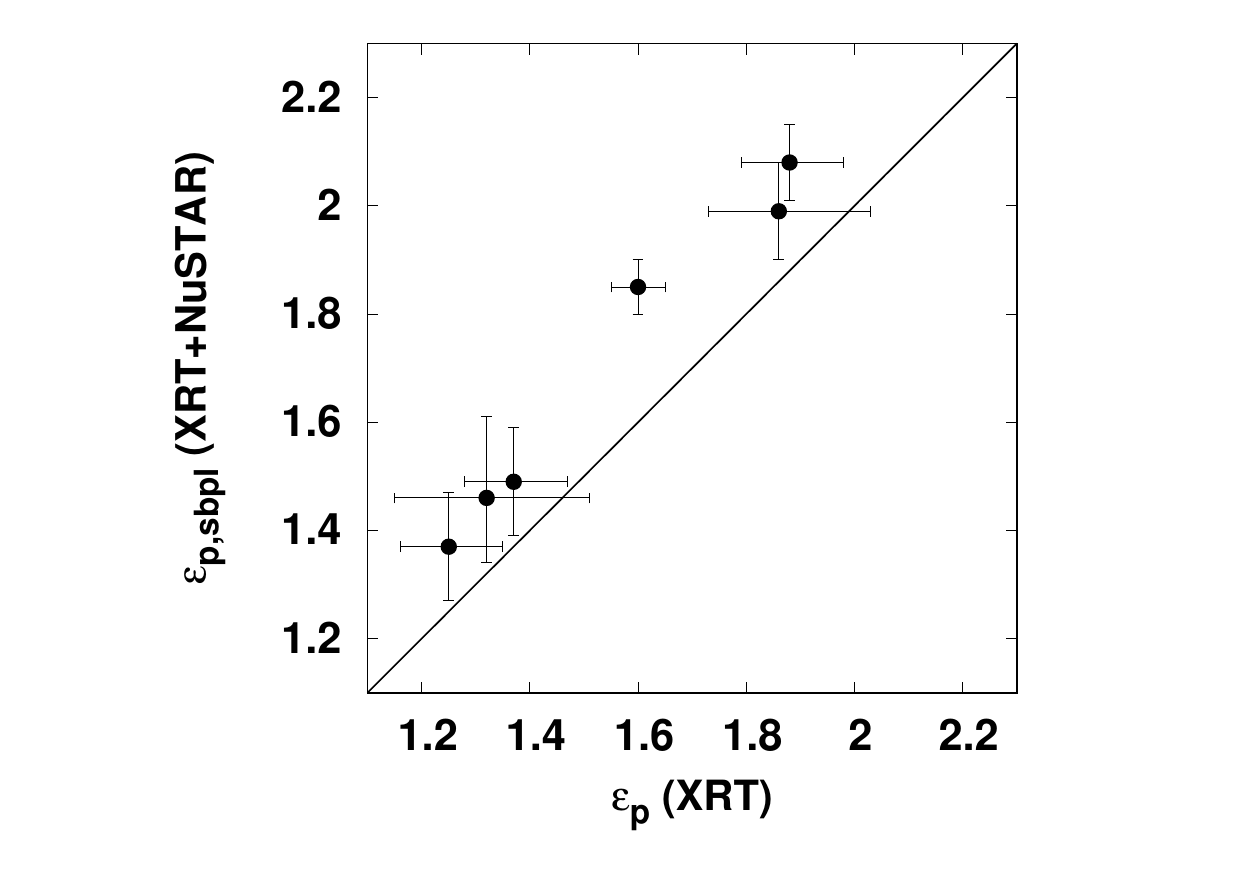}
\caption { The scatter plot between $\epsilon_{p}$ and $\epsilon_{p,sbpl}$ using combined \emph{Swift}-XRT and \emph{Nu}STAR data, along with the identity line.} 
\label{fig:logpar-ep-sbpl-ep-nustar}
\end{figure}

 An important uncertainty in the present work is related to the choice of $\epsilon_b$ used in spectral fitting 
with the \emph{sbpl} model. Since this parameter value has been frozen to $\epsilon_p$ obtained from the log 
parabola (symmetric function) spectral fit, this could introduce an additional bias which may be reflected in the 
parameters $\Gamma_{\rm high}$ and $\Gamma_{\rm low}$. In order to explore this, we performed a combined 
spectral fit using simultaneous/near-simultaneous \emph{Nu}STAR observations of the source.
The \emph{Nu}STAR data were downloaded from the online data archive\footnote{\url{https://heasarc.gsfc.nasa.gov/}}
and standard data reducing techniques were employed using the latest software provided\footnote{\url{https://heasarc.gsfc.nasa.gov/docs/nustar/analysis/}}.
The spectral fit is then performed using \emph{sbpl} model on simultaneous/near-simultaneous \emph{Swift}-XRT 
(0.3--10.0 keV) and \emph{Nu}STAR  (3.0 to 30.0 keV) X-ray spectra with all parameters set free. 
It may be noted that, only four \emph{Nu}STAR observations were simultaneous with \emph{Swift}-XRT; dividing the 
later orbit-wise we obtain six spectra. The best spectral fit parameters are given in Table \ref{tab:nustar-spectral-fitting}. 
We find that the $\epsilon_{\rm p,sbpl}$ obtained through the combined spectral fit matches reasonably well with  
$\epsilon_p$ got from the log parabolic fit of \emph{Swift}-XRT observations.
The scatter plot between these quantities along with the identity line is shown in Figure \ref{fig:logpar-ep-sbpl-ep-nustar}. 
Though this result can be viewed as supporting our choice of $\epsilon_b$, it is based on only a small number of data 
points. To improve on this and to be able to draw a more firm conclusion on the nature of $\epsilon_p$, further simultaneous
broad band X-ray observations of the source are clearly desirable. Nevertheless, the fact that all data points in Figure 
\ref{fig:theoretical-fit} lie above the cooling break line (brown dotted dash line) clearly disfavours a radiative loss 
interpretation of the considered, characteristic photon energy in Mkn\,421. 

A similar finding to the one presented here has also been reported by \cite{2011ApJ...736..131A} based on 
broadband SED modelling of the source observed in 2009. Their modelling results show that the required 
spectral break is significantly larger than the one inferred from a simple radiative cooling scenario. As a 
possible explanation the authors suggest that source inhomogeneities might be responsible for the large 
spectral break \citep[e.g.,][]{Reynolds_2009}.
Using a large number of \emph{Swift-XRT} observations corresponding to different flux states, we have
show here that a simple cooling break interpretation for the characteristic photon energy of Mkn\,421 is 
inconsistent with the observations. Our statistical analysis reveals a strong anti-correlation between 
$\Gamma_{\rm low}$ and $\Gamma_{\rm high}$, providing an important piece of information for identifying
the physics behind the characteristic photon energy. While simple scenarios can reproduce part of the 
observed properties, a deeper understanding of the origin and evolution of the radiating particle distribution 
is needed to satisfactorily match the observed results.

The correlation between the indices around the spectral peak might also be present for the high energy 
Compton spectral component if the same electron distribution is responsible for the emission at these 
energies and scattering occurs in the Thomson regime. However, the curvature could differ depending 
on the spectral shape of the target photons responsible for Compton emission. This could be further 
complicated if the Compton spectral peak is governed by the transition of the scattering process from the 
Thomson to the Klein-Nishina regime. 
We note that \cite{Chen_2014} has performed a detailed spectral study of a sample of 
\emph{Fermi} bright blazars by fitting their synchrotron and Compton spectral components (one SED data 
set for each source) log-parabola and broken power-law functions. Comparing the sources in this 
sample, a significant correlation between the peak frequency and the curvature was found for the synchrotron 
spectral components, while the Compton spectral components did not show such a behaviour. This 
seems at variance with our findings based on multiple observations of a single source where no 
such correlation has been seen for the synchrotron spectral component.
In principle, this could suggest that correlation results obtained from a single source cannot be easily 
generalised to the entire set of blazars.
On the other hand, the results for the Compton emission are still less certain given the (then) limited 
observational characterisation of the Compton spectrum, and an updated analysis might be helpful 
to clarify the situation. 

Earlier studies interpreting the blazar sequence in terms of radiative loss suggested FSRQs to undergo 
significant losses compared to BL Lacs  \citep[e.g.,][]{2002A&A...386..833G}. This results in a rather 
low characteristic (peak) photon energy for FSRQs compared to BL Lacs, though the latter are more 
luminous. However, the present study indicates that the characteristic photon energy is not readily 
related to radiative losses, at least in the case of Mkn 421, and hence this inference does not apply.
Therefore, a definite explanation of the characteristic photon energy along with a study based on an 
increase number of the sample will have the potential to decipher the mystery behind the blazar 
sequence.

\begin{table*}
\begin{center}
\caption{Best fit parameters of combined (\emph{Swift}-XRT and \emph{Nu}STAR) spectral fitting using \emph{sbpl} model} 
\label{tab:nustar-spectral-fitting}         
\begin{tabular}{c c c c c c c }
\hline

 \emph{Swift}-XRT ObsID 	&	\emph{Nu}STAR ObsID	& $\epsilon_{p,sbpl}$  & $\Gamma_{\rm low}$ & $\Gamma_{\rm high}$&  $\chi_{red}^2$ (dof) \\
\hline
 
 80050019-Orb1	&	60002023027	&	2.08$_{-0.07}^{+0.07}$	&	1.58$_{-0.03}^{+0.03}$	&	3.04$_{-0.03}^{+0.03}$	&			1.06 (1303)	\\
80050019-Orb5	&	60002023027	&				1.85$_{-0.05}^{+0.05}$	&	1.68	$_{-0.02}^{+0.02}$	&	3.08$_{-0.03}^{+0.03}$		&	1.05 (1438)	\\
35014065-Orb3	&	60002023035	&			1.99$_{-0.09}^{+0.09}$&	1.82$_{-0.02}^{+0.02}$	&	3.00$_{-0.05}^{+0.05}$			&	1.01 (1563)		\\
34228110-Orb3	&	60202048002	&			1.49$_{-0.1}^{+0.1}$		&	1.70$_{-0.07}^{+0.04}$	&	2.78$_{-0.07}^{+0.05}$			&	1.09 (1211)	\\
34228110-Orb5	&	60202048002	&				1.37$_{-0.1}^{+0.1}$		&	1.73$_{-0.07}^{+0.04}$	&	2.79$_{-0.07}^{+0.05}$&		1.04 (1201)	\\
81926001	&	60202048004	&			1.46$_{-0.12}^{+0.15}$		&	1.84$_{-0.04}^{+0.03}$		&	2.98$_{-0.08}^{+0.06}$&		1.03 (1240)	\\

\hline
\end{tabular}
\end{center}
\end{table*}

\begin{figure*}
    \centering
    \begin{subfigure}[h!]{0.45\linewidth}
        \includegraphics[height=84mm,width=75mm,angle=-90]{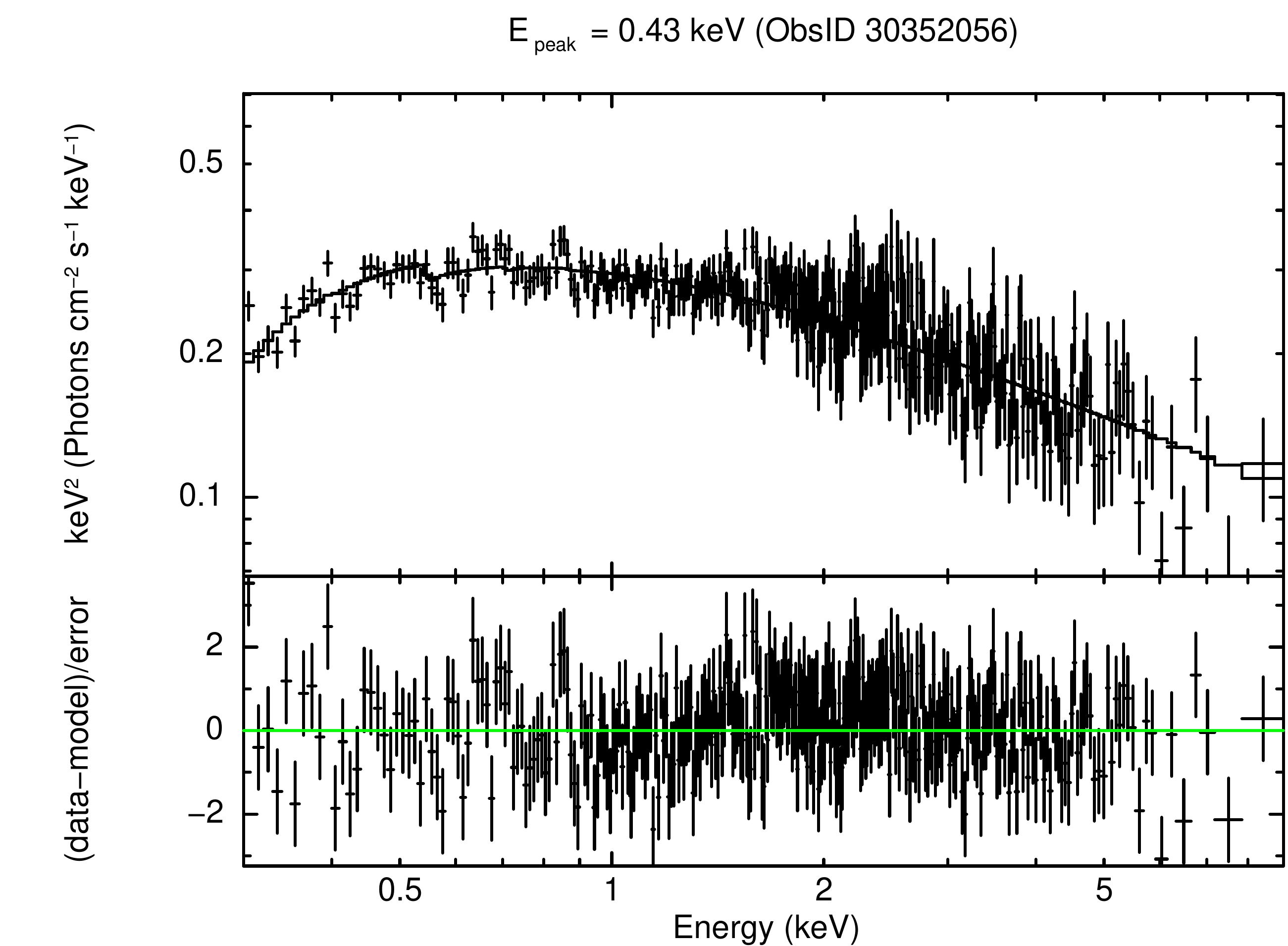}
    \end{subfigure}
        \vspace{2mm}
        \hspace{2mm}
    \begin{subfigure}[h!]{0.45\linewidth}
        \includegraphics[height=84mm,width=75mm,angle=-90]{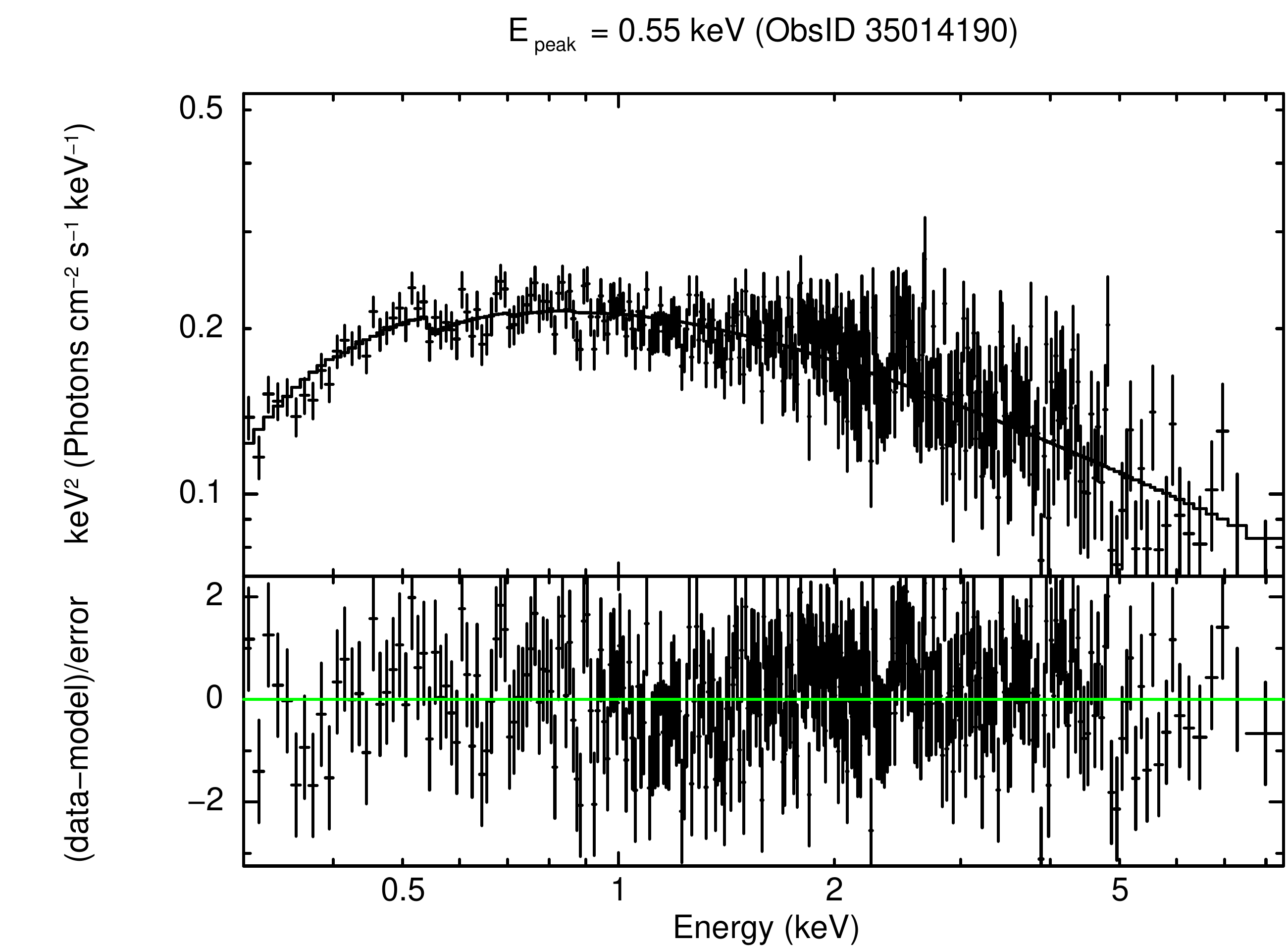}
    \end{subfigure}
    \vspace{2mm}
    \centering
    \hspace*{1mm}
    \begin{subfigure}[h!]{0.45\linewidth}
        \includegraphics[height=84mm,width=75mm,angle=-90]{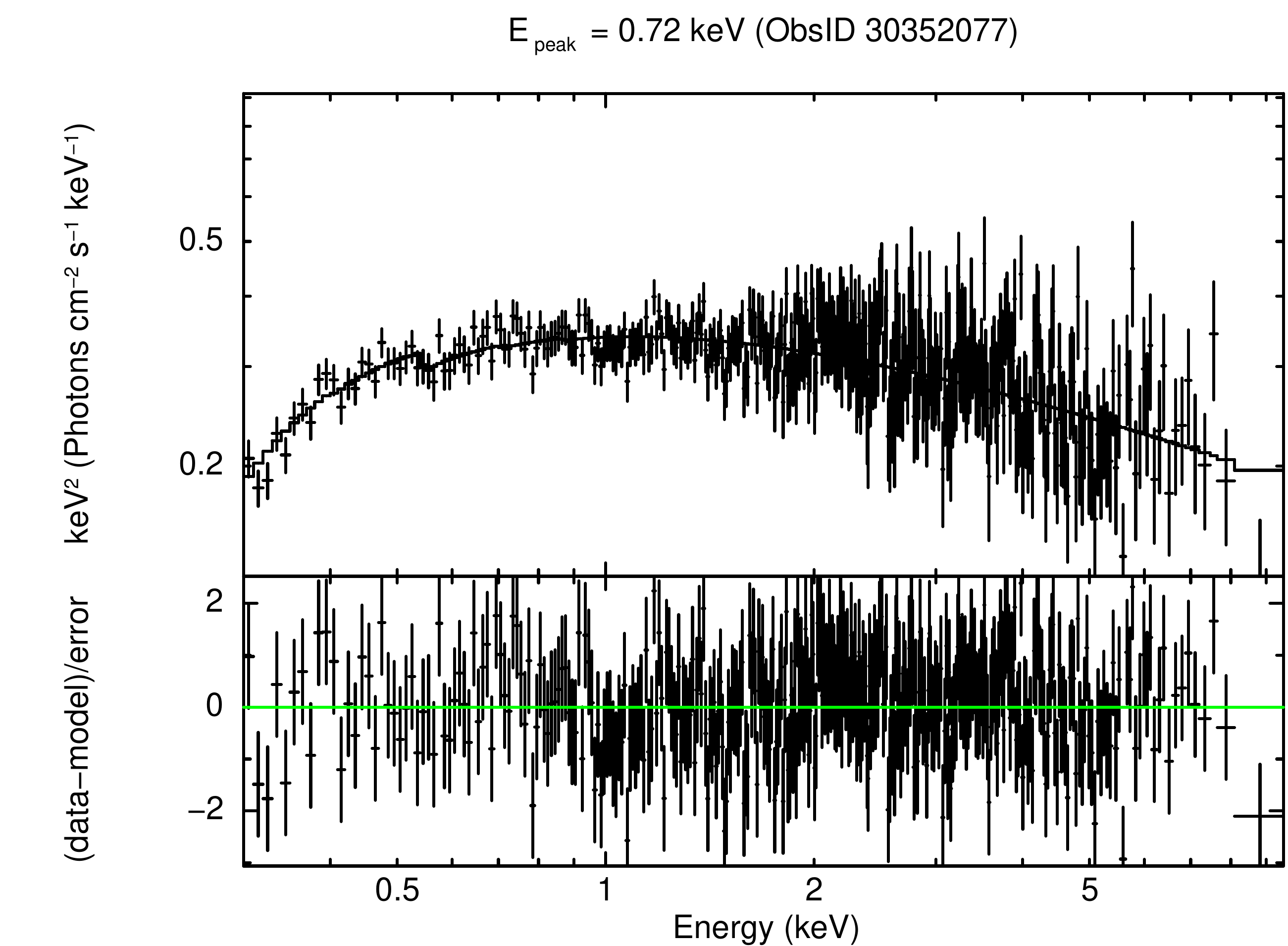}
    \end{subfigure}
        \vspace{2mm}
        \hspace*{1mm}
    \begin{subfigure}[h!]{0.45\linewidth}
        \includegraphics[height=84mm,width=75mm,angle=-90]{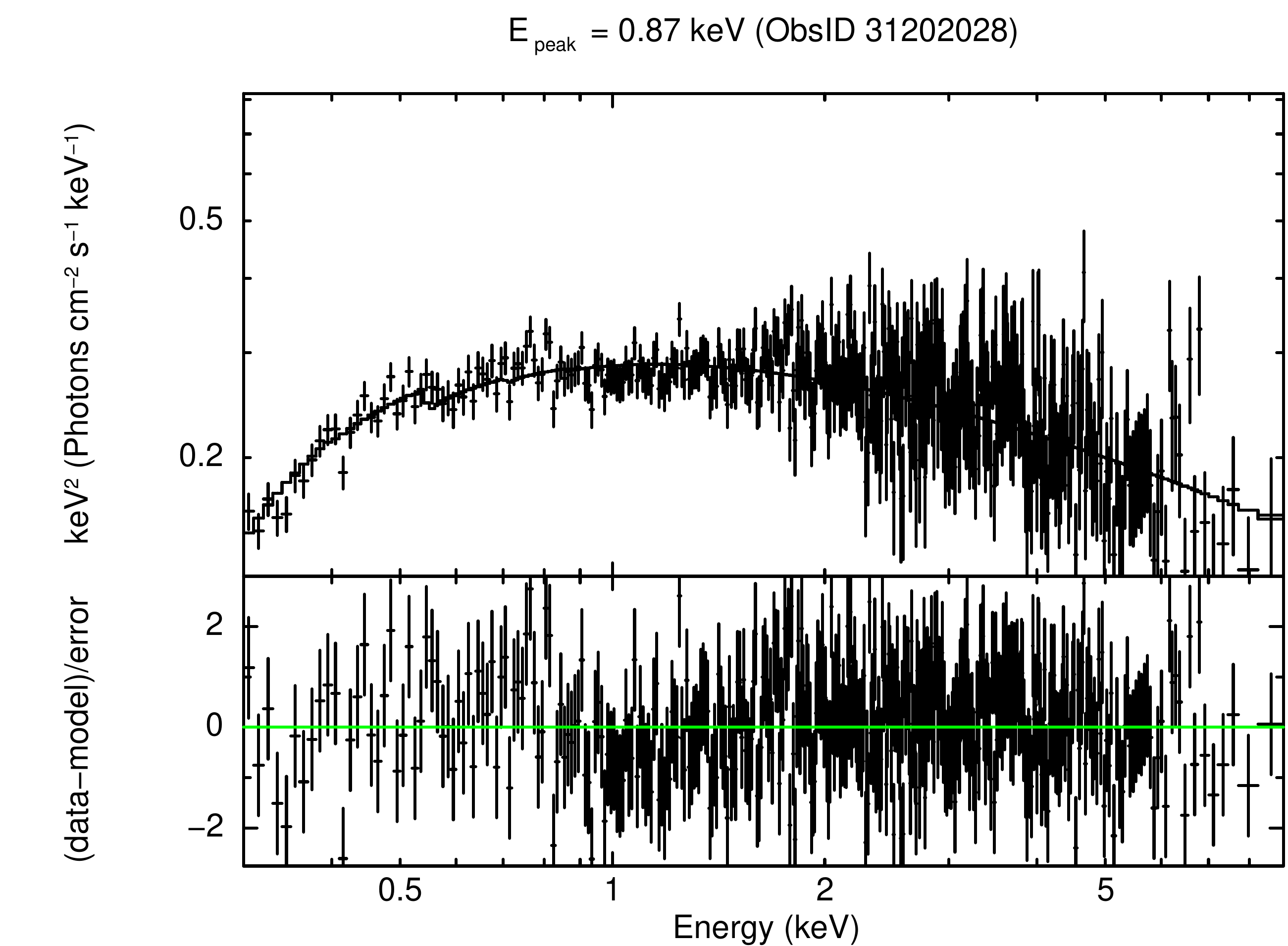}
    \end{subfigure}
    \hspace*{1mm}
    \centering
    \begin{subfigure}[h!]{0.45\linewidth}
        \includegraphics[height=84mm,width=75mm,angle=-90]{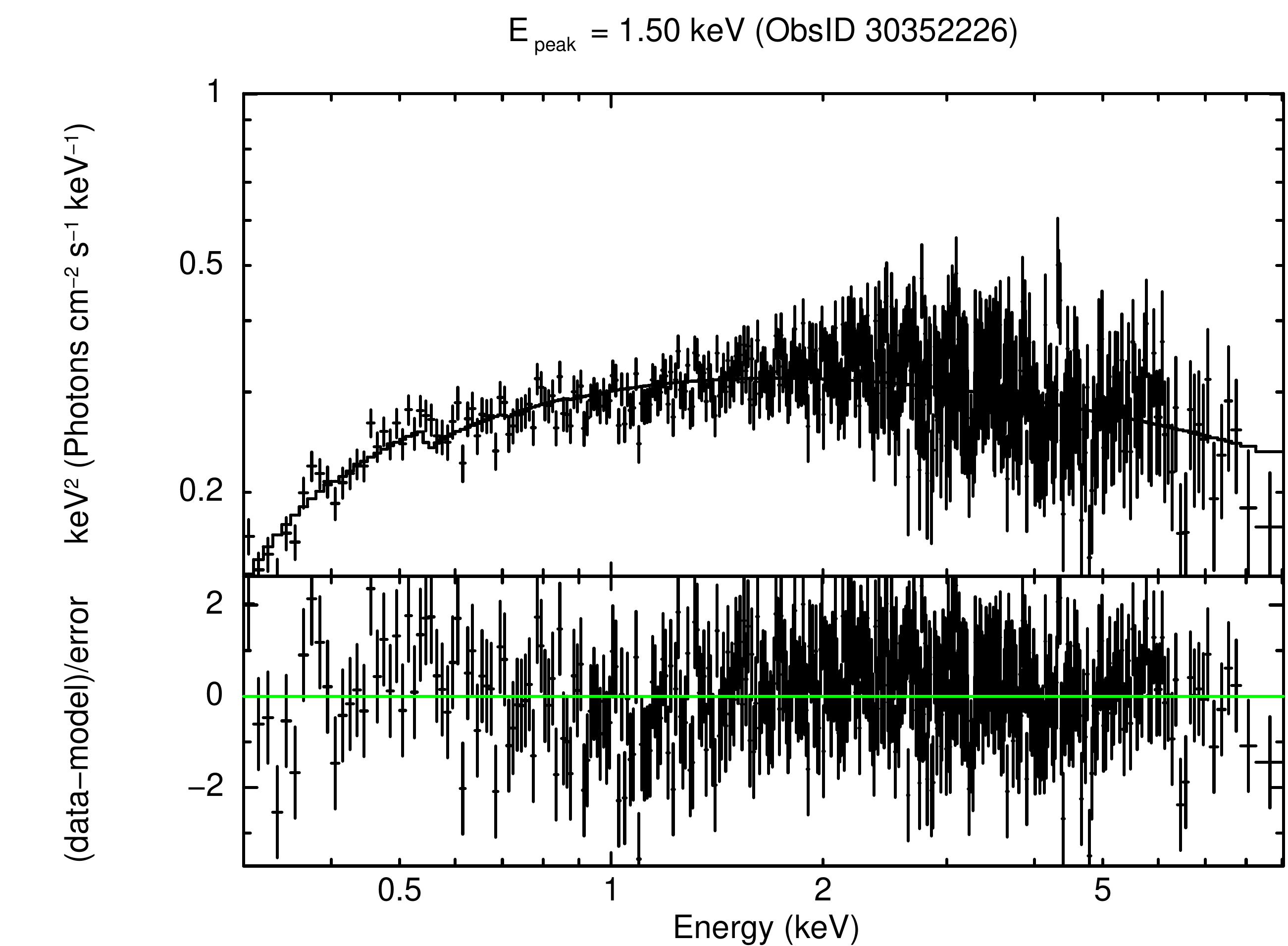}
    \end{subfigure}
        \vspace{2mm}
        \hspace*{1mm}
    \begin{subfigure}[h!]{0.45\linewidth}
        \includegraphics[height=84mm,width=75mm,angle=-90]{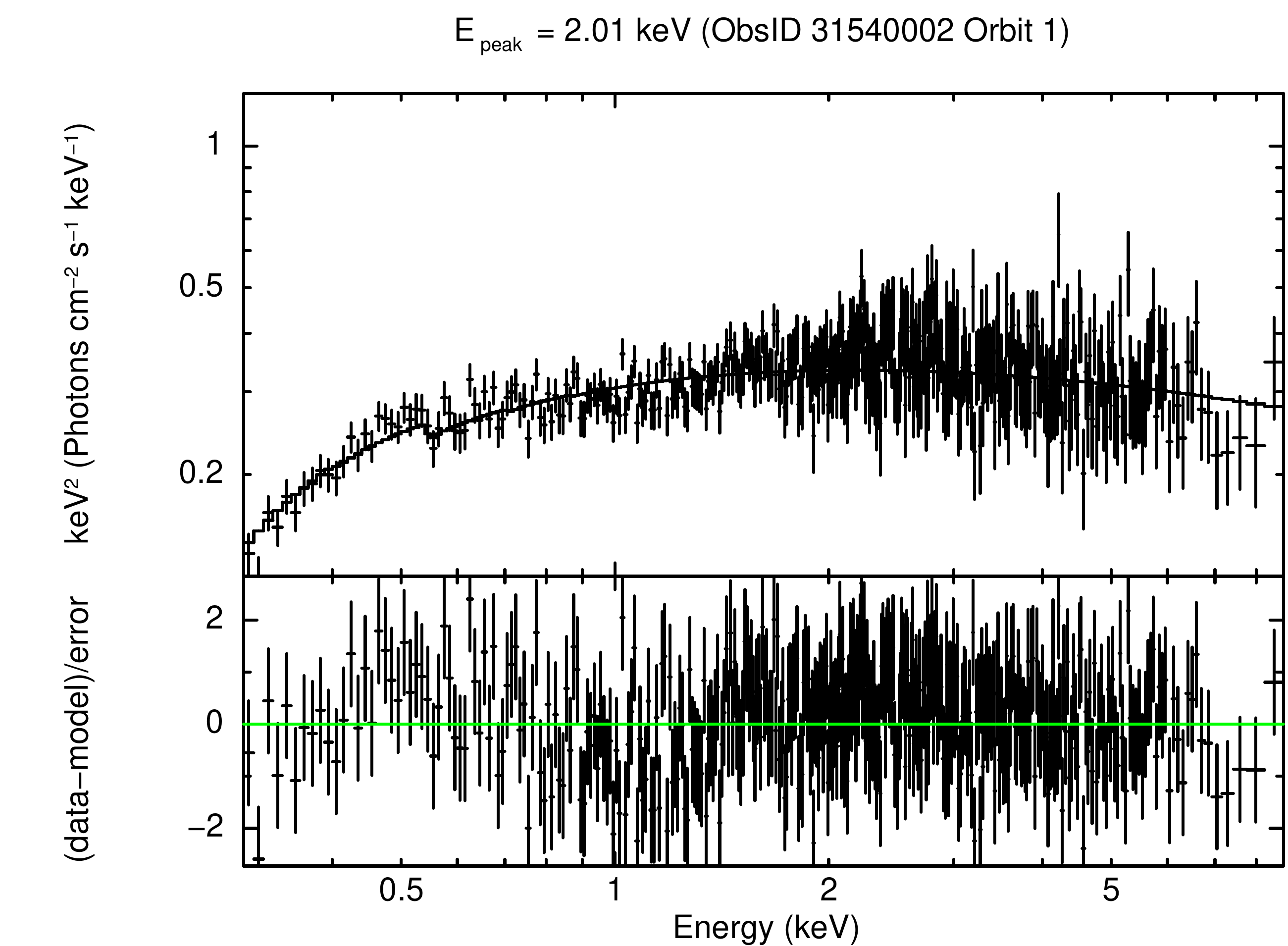}
    \end{subfigure}
        \caption{\emph{sbpl} fit for observation with different $\epsilon_p$ values.}
       \label{fig:sbpl-fit}
\end{figure*}

\section*{Acknowledgements}
We gratefully acknowledge the anonymous referee for very constructive comments. 
This research has made use of data obtained from NASA's High Energy Astrophysics 
Science Archive Research Center (HEASARC), a service of the Goddard Space Flight 
Center and the Smithsonian Astrophysical Observatory.
CB wishes to thank CSIR, New Delhi (09/043(0198)/2018-EMR-I) for financial support. 
CB is thankful to UGC-SAP and FIST 2 (SR/FIST/PS1-159/2010) (DST, Government of India) 
for the research facilities in the Department of Physics, University of Calicut.

\section*{Data Availability}

The data underlying this paper are publicly available from the
HEASARC archive at \url{https://heasarc.gsfc.nasa.gov/}.




\bibliographystyle{mnras}
\bibliography{ms} 

\begin{thebibliography}{}
\makeatletter
\relax
\def\mn@urlcharsother{\let\do\@makeother \do\$\do\&\do\#\do\^\do\_\do\%\do\~}
\def\mn@doi{\begingroup\mn@urlcharsother \@ifnextchar [ {\mn@doi@}
  {\mn@doi@[]}}
\def\mn@doi@[#1]#2{\def\@tempa{#1}\ifx\@tempa\@empty \href
  {http://dx.doi.org/#2} {doi:#2}\else \href {http://dx.doi.org/#2} {#1}\fi
  \endgroup}
\def\mn@eprint#1#2{\mn@eprint@#1:#2::\@nil}
\def\mn@eprint@arXiv#1{\href {http://arxiv.org/abs/#1} {{\tt arXiv:#1}}}
\def\mn@eprint@dblp#1{\href {http://dblp.uni-trier.de/rec/bibtex/#1.xml}
  {dblp:#1}}
\def\mn@eprint@#1:#2:#3:#4\@nil{\def\@tempa {#1}\def\@tempb {#2}\def\@tempc
  {#3}\ifx \@tempc \@empty \let \@tempc \@tempb \let \@tempb \@tempa \fi \ifx
  \@tempb \@empty \def\@tempb {arXiv}\fi \@ifundefined
  {mn@eprint@\@tempb}{\@tempb:\@tempc}{\expandafter \expandafter \csname
  mn@eprint@\@tempb\endcsname \expandafter{\@tempc}}}

\bibitem[\protect\citeauthoryear{{Abdo} et~al.,}{{Abdo}
  et~al.}{2011}]{2011ApJ...736..131A}
{Abdo} A.~A.,  et~al., 2011, \mn@doi [\apj] {10.1088/0004-637X/736/2/131},
  \href {https://ui.adsabs.harvard.edu/abs/2011ApJ...736..131A} {736, 131}

\bibitem[\protect\citeauthoryear{{Acciari} et~al.,}{{Acciari}
  et~al.}{2009}]{2009ApJ...703..169A}
{Acciari} V.~A.,  et~al., 2009, \mn@doi [\apj] {10.1088/0004-637X/703/1/169},
  \href {https://ui.adsabs.harvard.edu/abs/2009ApJ...703..169A} {703, 169}

\bibitem[\protect\citeauthoryear{{Acciari} et~al.,}{{Acciari}
  et~al.}{2011}]{2011ApJ...738...25A}
{Acciari} V.~A.,  et~al., 2011, \mn@doi [\apj] {10.1088/0004-637X/738/1/25},
  \href {https://ui.adsabs.harvard.edu/abs/2011ApJ...738...25A} {738, 25}

\bibitem[\protect\citeauthoryear{{Aleksi{\'c}} et~al.,}{{Aleksi{\'c}}
  et~al.}{2012}]{2012A&A...542A.100A}
{Aleksi{\'c}} J.,  et~al., 2012, \mn@doi [\aap] {10.1051/0004-6361/201117442},
  \href {https://ui.adsabs.harvard.edu/abs/2012A&A...542A.100A} {542, A100}

\bibitem[\protect\citeauthoryear{{Arnaud}}{{Arnaud}}{1996}]{1996ASPC..101...17A}
{Arnaud} K.~A.,  1996, in {Jacoby} G.~H.,  {Barnes} J.,  eds,  Astronomical
  Society of the Pacific Conference Series Vol. 101, Astronomical Data Analysis
  Software and Systems V. p.~17

\bibitem[\protect\citeauthoryear{{Balokovi{\'c}} et~al.,}{{Balokovi{\'c}}
  et~al.}{2016}]{2016ApJ...819..156B}
{Balokovi{\'c}} M.,  et~al., 2016, \mn@doi [\apj]
  {10.3847/0004-637X/819/2/156}, \href
  {https://ui.adsabs.harvard.edu/abs/2016ApJ...819..156B} {819, 156}

\bibitem[\protect\citeauthoryear{{Bartoli} et~al.,}{{Bartoli}
  et~al.}{2011}]{2011ApJ...734..110B}
{Bartoli} B.,  et~al., 2011, \mn@doi [\apj] {10.1088/0004-637X/734/2/110},
  \href {https://ui.adsabs.harvard.edu/abs/2011ApJ...734..110B} {734, 110}

\bibitem[\protect\citeauthoryear{{Bartoli} et~al.,}{{Bartoli}
  et~al.}{2016}]{2016ApJS..222....6B}
{Bartoli} B.,  et~al., 2016, \mn@doi [\apjs] {10.3847/0067-0049/222/1/6}, \href
  {https://ui.adsabs.harvard.edu/abs/2016ApJS..222....6B} {222, 6}

\bibitem[\protect\citeauthoryear{Begelman, Blandford  \& Rees}{Begelman
  et~al.}{1984}]{RevModPhys.56.255}
Begelman M.~C.,  Blandford R.~D.,   Rees M.~J.,  1984, \mn@doi [Rev. Mod.
  Phys.] {10.1103/RevModPhys.56.255}, 56, 255

\bibitem[\protect\citeauthoryear{{B{\l}a{\.z}ejowski}
  et~al.,}{{B{\l}a{\.z}ejowski} et~al.}{2005}]{2005ApJ...630..130B}
{B{\l}a{\.z}ejowski} M.,  et~al., 2005, \mn@doi [\apj] {10.1086/431925}, \href
  {https://ui.adsabs.harvard.edu/abs/2005ApJ...630..130B} {630, 130}

\bibitem[\protect\citeauthoryear{{Burrows} et~al.,}{{Burrows}
  et~al.}{2005}]{2005SSRv..120..165B}
{Burrows} D.~N.,  et~al., 2005, \mn@doi [\ssr] {10.1007/s11214-005-5097-2},
  \href {https://ui.adsabs.harvard.edu/abs/2005SSRv..120..165B} {120, 165}

\bibitem[\protect\citeauthoryear{Celotti \& Ghisellini}{Celotti \&
  Ghisellini}{2008}]{10.1111/j.1365-2966.2007.12758.x}
Celotti A.,  Ghisellini G.,  2008, \mn@doi [Monthly Notices of the Royal
  Astronomical Society] {10.1111/j.1365-2966.2007.12758.x}, 385, 283

\bibitem[\protect\citeauthoryear{Chen}{Chen}{2014}]{Chen_2014}
Chen L.,  2014, \mn@doi [The Astrophysical Journal]
  {10.1088/0004-637x/788/2/179}, 788, 179

\bibitem[\protect\citeauthoryear{{Dondi} \& {Ghisellini}}{{Dondi} \&
  {Ghisellini}}{1995}]{1995MNRAS.273..583D}
{Dondi} L.,  {Ghisellini} G.,  1995, \mn@doi [\mnras]
  {10.1093/mnras/273.3.583}, \href
  {https://ui.adsabs.harvard.edu/abs/1995MNRAS.273..583D} {273, 583}

\bibitem[\protect\citeauthoryear{{Fossati} et~al.,}{{Fossati}
  et~al.}{2008}]{2008ApJ...677..906F}
{Fossati} G.,  et~al., 2008, \mn@doi [\apj] {10.1086/527311}, \href
  {https://ui.adsabs.harvard.edu/abs/2008ApJ...677..906F} {677, 906}

\bibitem[\protect\citeauthoryear{{Ghisellini}, {Celotti}  \&
  {Costamante}}{{Ghisellini} et~al.}{2002}]{2002A&A...386..833G}
{Ghisellini} G.,  {Celotti} A.,   {Costamante} L.,  2002, \mn@doi [\aap]
  {10.1051/0004-6361:20020275}, \href
  {https://ui.adsabs.harvard.edu/abs/2002A&A...386..833G} {386, 833}

\bibitem[\protect\citeauthoryear{{Giebels}, {Dubus}  \&
  {Kh{\'e}lifi}}{{Giebels} et~al.}{2007}]{2007A&A...462...29G}
{Giebels} B.,  {Dubus} G.,   {Kh{\'e}lifi} B.,  2007, \mn@doi [\aap]
  {10.1051/0004-6361:20066134}, \href
  {https://ui.adsabs.harvard.edu/abs/2007A&A...462...29G} {462, 29}

\bibitem[\protect\citeauthoryear{Goswami, Sahayanathan, Sinha, Misra  \&
  Gogoi}{Goswami et~al.}{2018}]{10.1093/mnras/sty2003}
Goswami P.,  Sahayanathan S.,  Sinha A.,  Misra R.,   Gogoi R.,  2018, \mn@doi
  [Monthly Notices of the Royal Astronomical Society] {10.1093/mnras/sty2003},
  480, 2046

\bibitem[\protect\citeauthoryear{{Goswami}, {Sahayanathan}, {Sinha}  \&
  {Gogoi}}{{Goswami} et~al.}{2020}]{2020MNRAS.499.2094G}
{Goswami} P.,  {Sahayanathan} S.,  {Sinha} A.,   {Gogoi} R.,  2020, \mn@doi
  [\mnras] {10.1093/mnras/staa3022}, \href
  {https://ui.adsabs.harvard.edu/abs/2020MNRAS.499.2094G} {499, 2094}

\bibitem[\protect\citeauthoryear{{Jagan}, {Sahayanathan}, {Misra}, {Ravikumar}
  \& {Jeena}}{{Jagan} et~al.}{2018}]{2018MNRAS.478L.105J}
{Jagan} S.~K.,  {Sahayanathan} S.,  {Misra} R.,  {Ravikumar} C.~D.,   {Jeena}
  K.,  2018, \mn@doi [\mnras] {10.1093/mnrasl/sly086}, \href
  {https://ui.adsabs.harvard.edu/abs/2018MNRAS.478L.105J} {478, L105}

\bibitem[\protect\citeauthoryear{{Kalberla}, {Burton}, {Hartmann}, {Arnal},
  {Bajaja}, {Morras}  \& {P{\"o}ppel}}{{Kalberla}
  et~al.}{2005}]{2005A&A...440..775K}
{Kalberla} P.~M.~W.,  {Burton} W.~B.,  {Hartmann} D.,  {Arnal} E.~M.,  {Bajaja}
  E.,  {Morras} R.,   {P{\"o}ppel} W.~G.~L.,  2005, \mn@doi [\aap]
  {10.1051/0004-6361:20041864}, \href
  {https://ui.adsabs.harvard.edu/abs/2005A&A...440..775K} {440, 775}

\bibitem[\protect\citeauthoryear{{Kapanadze} et~al.,}{{Kapanadze}
  et~al.}{2016}]{2016ApJ...831..102K}
{Kapanadze} B.,  et~al., 2016, \mn@doi [\apj] {10.3847/0004-637X/831/1/102},
  \href {https://ui.adsabs.harvard.edu/abs/2016ApJ...831..102K} {831, 102}

\bibitem[\protect\citeauthoryear{Kapanadze, Dorner, Romano, Vercellone,
  Kapanadze  \& Tabagari}{Kapanadze et~al.}{2017}]{Kapanadze_2017}
Kapanadze B.,  Dorner D.,  Romano P.,  Vercellone S.,  Kapanadze S.,   Tabagari
  L.,  2017, \mn@doi [The Astrophysical Journal] {10.3847/1538-4357/aa8ea6},
  848, 103

\bibitem[\protect\citeauthoryear{{Kapanadze} et~al.,}{{Kapanadze}
  et~al.}{2018a}]{2018ApJ...854...66K}
{Kapanadze} B.,  et~al., 2018a, \mn@doi [\apj] {10.3847/1538-4357/aaa75d},
  \href {https://ui.adsabs.harvard.edu/abs/2018ApJ...854...66K} {854, 66}

\bibitem[\protect\citeauthoryear{{Kapanadze}, {Vercellone}, {Romano}, {Hughes},
  {Aller}, {Aller}, {Kharshiladze}  \& {Tabagari}}{{Kapanadze}
  et~al.}{2018b}]{2018ApJ...858...68K}
{Kapanadze} B.,  {Vercellone} S.,  {Romano} P.,  {Hughes} P.,  {Aller} M.,
  {Aller} H.,  {Kharshiladze} O.,   {Tabagari} L.,  2018b, \mn@doi [\apj]
  {10.3847/1538-4357/aabbac}, \href
  {https://ui.adsabs.harvard.edu/abs/2018ApJ...858...68K} {858, 68}

\bibitem[\protect\citeauthoryear{Kapanadze et~al.,}{Kapanadze
  et~al.}{2020}]{Kapanadze_2020}
Kapanadze B.,  et~al., 2020, \mn@doi [The Astrophysical Journal Supplement
  Series] {10.3847/1538-4365/ab6322}, 247, 27

\bibitem[\protect\citeauthoryear{{Kardashev}}{{Kardashev}}{1962}]{1962SvA.....6..317K}
{Kardashev} N.~S.,  1962, \sovast, \href
  {https://ui.adsabs.harvard.edu/abs/1962SvA.....6..317K} {6, 317}

\bibitem[\protect\citeauthoryear{{Kataoka} \& {Stawarz}}{{Kataoka} \&
  {Stawarz}}{2016}]{2016ApJ...827...55K}
{Kataoka} J.,  {Stawarz} {\L}.,  2016, \mn@doi [\apj]
  {10.3847/0004-637X/827/1/55}, \href
  {https://ui.adsabs.harvard.edu/abs/2016ApJ...827...55K} {827, 55}

\bibitem[\protect\citeauthoryear{{Katarzy{\'n}ski} \&
  {Walczewska}}{{Katarzy{\'n}ski} \& {Walczewska}}{2010}]{2010A&A...510A..63K}
{Katarzy{\'n}ski} K.,  {Walczewska} K.,  2010, \mn@doi [\aap]
  {10.1051/0004-6361/200913128}, \href
  {https://ui.adsabs.harvard.edu/abs/2010A&A...510A..63K} {510, A63}

\bibitem[\protect\citeauthoryear{{Kirk}, {Rieger}  \& {Mastichiadis}}{{Kirk}
  et~al.}{1998}]{1998A&A...333..452K}
{Kirk} J.~G.,  {Rieger} F.~M.,   {Mastichiadis} A.,  1998, \aap, \href
  {https://ui.adsabs.harvard.edu/abs/1998A&A...333..452K} {333, 452}

\bibitem[\protect\citeauthoryear{Krawczynski et~al.,}{Krawczynski
  et~al.}{2004}]{Krawczynski_2004}
Krawczynski H.,  et~al., 2004, \mn@doi [The Astrophysical Journal]
  {10.1086/380393}, 601, 151

\bibitem[\protect\citeauthoryear{Mankuzhiyil, Ansoldi, Persic, Rivers,
  Rothschild  \& Tavecchio}{Mankuzhiyil et~al.}{2012}]{Mankuzhiyil_2012}
Mankuzhiyil N.,  Ansoldi S.,  Persic M.,  Rivers E.,  Rothschild R.,
  Tavecchio F.,  2012, \mn@doi [The Astrophysical Journal]
  {10.1088/0004-637x/753/2/154}, 753, 154

\bibitem[\protect\citeauthoryear{{Massaro}, {Perri}, {Giommi}  \&
  {Nesci}}{{Massaro} et~al.}{2004}]{2004A&A...413..489M}
{Massaro} E.,  {Perri} M.,  {Giommi} P.,   {Nesci} R.,  2004, \mn@doi [\aap]
  {10.1051/0004-6361:20031558}, \href
  {https://ui.adsabs.harvard.edu/abs/2004A&A...413..489M} {413, 489}

\bibitem[\protect\citeauthoryear{{Massaro}, {Tramacere}, {Cavaliere}, {Perri}
  \& {Giommi}}{{Massaro} et~al.}{2008}]{2008A&A...478..395M}
{Massaro} F.,  {Tramacere} A.,  {Cavaliere} A.,  {Perri} M.,   {Giommi} P.,
  2008, \mn@doi [\aap] {10.1051/0004-6361:20078639}, \href
  {https://ui.adsabs.harvard.edu/abs/2008A&A...478..395M} {478, 395}

\bibitem[\protect\citeauthoryear{{Mastichiadis} \& {Kirk}}{{Mastichiadis} \&
  {Kirk}}{1997}]{1997A&A...320...19M}
{Mastichiadis} A.,  {Kirk} J.~G.,  1997, \aap, \href
  {https://ui.adsabs.harvard.edu/abs/1997A&A...320...19M} {320, 19}

\bibitem[\protect\citeauthoryear{{Press}, {Teukolsky}, {Vetterling}  \&
  {Flannery}}{{Press} et~al.}{1992}]{1992nrfa.book.....P}
{Press} W.~H.,  {Teukolsky} S.~A.,  {Vetterling} W.~T.,   {Flannery} B.~P.,
  1992, {Numerical recipes in FORTRAN. The art of scientific computing}

\bibitem[\protect\citeauthoryear{Reynolds}{Reynolds}{2009}]{Reynolds_2009}
Reynolds S.~P.,  2009, \mn@doi [The Astrophysical Journal]
  {10.1088/0004-637x/703/1/662}, 703, 662

\bibitem[\protect\citeauthoryear{{Rieger}, {Bosch-Ramon}  \& {Duffy}}{{Rieger}
  et~al.}{2007}]{2007Ap&SS.309..119R}
{Rieger} F.~M.,  {Bosch-Ramon} V.,   {Duffy} P.,  2007, \mn@doi [\apss]
  {10.1007/s10509-007-9466-z}, \href
  {https://ui.adsabs.harvard.edu/abs/2007Ap&SS.309..119R} {309, 119}

\bibitem[\protect\citeauthoryear{{Rybicki} \& {Lightman}}{{Rybicki} \&
  {Lightman}}{1986}]{1986rpa..book.....R}
{Rybicki} G.~B.,  {Lightman} A.~P.,  1986, {Radiative Processes in
  Astrophysics}

\bibitem[\protect\citeauthoryear{{Sahayanathan}}{{Sahayanathan}}{2008}]{2008MNRAS.388L..49S}
{Sahayanathan} S.,  2008, \mn@doi [\mnras] {10.1111/j.1745-3933.2008.00497.x},
  \href {https://ui.adsabs.harvard.edu/abs/2008MNRAS.388L..49S} {388, L49}

\bibitem[\protect\citeauthoryear{{Sinha}, {Shukla}, {Misra}, {Chitnis}, {Rao}
  \& {Acharya}}{{Sinha} et~al.}{2015a}]{2015A&A...580A.100S}
{Sinha} A.,  {Shukla} A.,  {Misra} R.,  {Chitnis} V.~R.,  {Rao} A.~R.,
  {Acharya} B.~S.,  2015a, \mn@doi [\aap] {10.1051/0004-6361/201526264}, \href
  {https://ui.adsabs.harvard.edu/abs/2015A&A...580A.100S} {580, A100}

\bibitem[\protect\citeauthoryear{Sinha, Shukla, Misra, Chitnis, Rao  \&
  Acharya}{Sinha et~al.}{2015b}]{Sinha_2015}
Sinha A.,  Shukla A.,  Misra R.,  Chitnis V.~R.,  Rao A.~R.,   Acharya B.~S.,
  2015b, \mn@doi [Astronomy & Astrophysics] {10.1051/0004-6361/201526264}, 580,
  A100

\bibitem[\protect\citeauthoryear{{Sinha}, {Sahayanathan}, {Acharya}, {Anupama},
  {Chitnis}  \& {Singh}}{{Sinha} et~al.}{2017}]{2017ApJ...836...83S}
{Sinha} A.,  {Sahayanathan} S.,  {Acharya} B.~S.,  {Anupama} G.~C.,  {Chitnis}
  V.~R.,   {Singh} B.~B.,  2017, \mn@doi [\apj] {10.3847/1538-4357/836/1/83},
  \href {https://ui.adsabs.harvard.edu/abs/2017ApJ...836...83S} {836, 83}

\bibitem[\protect\citeauthoryear{{Tramacere}, {Massaro}  \&
  {Cavaliere}}{{Tramacere} et~al.}{2007a}]{2007A&A...466..521T}
{Tramacere} A.,  {Massaro} F.,   {Cavaliere} A.,  2007a, \mn@doi [\aap]
  {10.1051/0004-6361:20066723}, \href
  {https://ui.adsabs.harvard.edu/abs/2007A&A...466..521T} {466, 521}

\bibitem[\protect\citeauthoryear{{Tramacere} et~al.,}{{Tramacere}
  et~al.}{2007b}]{2007A&A...467..501T}
{Tramacere} A.,  et~al., 2007b, \mn@doi [\aap] {10.1051/0004-6361:20066226},
  \href {https://ui.adsabs.harvard.edu/abs/2007A&A...467..501T} {467, 501}

\bibitem[\protect\citeauthoryear{{Tramacere}, {Giommi}, {Perri}, {Verrecchia}
  \& {Tosti}}{{Tramacere} et~al.}{2009}]{2009A&A...501..879T}
{Tramacere} A.,  {Giommi} P.,  {Perri} M.,  {Verrecchia} F.,   {Tosti} G.,
  2009, \mn@doi [\aap] {10.1051/0004-6361/200810865}, \href
  {https://ui.adsabs.harvard.edu/abs/2009A&A...501..879T} {501, 879}

\bibitem[\protect\citeauthoryear{{Tramacere}, {Massaro}  \&
  {Taylor}}{{Tramacere} et~al.}{2011}]{2011ApJ...739...66T}
{Tramacere} A.,  {Massaro} E.,   {Taylor} A.~M.,  2011, \mn@doi [\apj]
  {10.1088/0004-637X/739/2/66}, \href
  {https://ui.adsabs.harvard.edu/abs/2011ApJ...739...66T} {739, 66}

\bibitem[\protect\citeauthoryear{Yan, Zhang, Yuan, Fan  \& Zeng}{Yan
  et~al.}{2013}]{Yan_2013}
Yan D.,  Zhang L.,  Yuan Q.,  Fan Z.,   Zeng H.,  2013, \mn@doi [The
  Astrophysical Journal] {10.1088/0004-637x/765/2/122}, 765, 122

\makeatother
\end{thebibliography}








\bsp	
\label{lastpage}

\end{document}


\begin{table*}
\begin{center}
\caption{Best fit parameters of spectral fitting using \emph{eplogpar} model and  \emph{sbpl} model} 
\label{tab:spectral-fitting}         
\begin{tabular}{l c c c c c c c c c }
\hline
 ObsID	&	Date of Obs.	&	Exposure	&		&	\emph{eplogpar}	&		&		&	\emph{sbpl}	&	 & 	Flux$_{0.3-10.0\,\rm kev}$\\
 
 \cline{4-6}
 \cline{7-9}
 
  	&		&	(sec)	&	$\epsilon_p$\,(keV)	&	$\beta$	&	$\tilde{\chi_{red}^2}$\,(dof)	&	$\Gamma_{\rm low}$	&	$\Gamma_{\rm high}$		&	$\tilde{\chi_{red}^2}$\,(dof) & ( $10^{-10} \rm erg\,cm^{-2} s^{-1}$)	\\
 
(1)	&	(2)	&	(3)	&	(4)	&	(5)	&	(6)	&	(7)		&	(8) & (9)&(10)	\\
\hline
\hline

30352053-Orb2	&	2008-01-16	&	575.1	&	0.42$^{+0.07}_{-0.08}$	&	0.28$\pm$0.05	&	1.04 (305)		&	1.36$\pm$0.12	&	2.54$\pm$0.02	&	1.06 (305)	&	20$\pm$0.28	\\
30352054	&	2008-01-16	&	1134.085	&	0.42$^{+0.04}_{-0.05}$	&	0.3$\pm$0.03	&	1.08 (394)		&	1.33$\pm$0.1	&	2.57$\pm$0.01	&	1.12 (394)	&	15.26$\pm$0.15	\\
30352056	&	2008-01-17	&	944.126	&	0.43$^{+0.05}_{-0.07}$	&	0.34$\pm$0.04	&	0.95 (322)		&	1.29$\pm$0.12	&	2.61$\pm$0.02	&	1 (322)	&	13.54$\pm$0.17	\\
30352055-Orb1	&	2008-01-17	&	394.9	&	0.64$^{+0.07}_{-0.08}$	&	0.45$\pm$0.08	&	1.11 (227)		&	1.29$\pm$0.12	&	2.69$\pm$0.04	&	1.12 (227)	&	10$\pm$0.21	\\
30352055-Orb2	&	2008-01-17	&	362.6	&	0.60$^{+0.05}_{-0.04}$	&	0.49$\pm$0.06	&	1.09 (260)		&	1.22$\pm$0.1	&	2.73$\pm$0.03	&	1.12 (260)	&	15.04$\pm$0.26	\\
30352058	&	2008-01-18	&	889.108	&	0.57$^{+0.04}_{-0.04}$	&	0.33$\pm$0.04	&	1.07 (378)		&	1.38$\pm$0.07	&	2.59$\pm$0.02	&	1.1 (378)	&	14.88$\pm$0.16	\\
30352059	&	2008-01-19	&	919.112	&	0.43$^{+0.04}_{-0.04}$	&	0.34$\pm$0.04	&	1.04 (359)		&	1.22$\pm$0.12	&	2.62$\pm$0.01	&	1.06 (359)	&	13.73$\pm$0.15	\\
30352060	&	2008-02-06	&	753.06	&	0.60$^{+0.05}_{-0.06}$	&	0.26$\pm$0.03	&	1.13 (429)		&	1.48$\pm$0.06	&	2.51$\pm$0.02	&	1.15 (429)	&	24.33$\pm$0.24	\\
30352066	&	2008-02-10	&	1174.1	&	0.74$^{+0.03}_{-0.03}$	&	0.3$\pm$0.03	&	1.17 (467)		&	1.46$\pm$0.04	&	2.55$\pm$0.02	&	1.18 (467)	&	20.85$\pm$0.18	\\
30352068	&	2008-02-11	&	1868.208	&	0.46$^{+0.03}_{-0.05}$	&	0.25$\pm$0.02	&	1.1 (495)		&	1.44$\pm$0.06	&	2.52$\pm$0.01	&	1.14 (495)	&	18.61$\pm$0.14	\\
30352069	&	2008-02-11	&	949.085	&	0.48$^{+0.04}_{-0.06}$	&	0.28$\pm$0.03	&	0.96 (422)		&	1.4$\pm$0.08	&	2.54$\pm$0.01	&	0.98 (422)	&	18.53$\pm$0.18	\\
30352071-Seg1	&	2008-02-12	&	699.7	&	0.5$^{+0.08}_{-0.07}$	&	0.24$\pm$0.04	&	1.14 (381)		&	1.49$\pm$0.09	&	2.49$\pm$0.02	&	1.17 (381)	&	18.9$\pm$0.22	\\
30352071-Seg2	&	2008-02-12	&	763.7	&	0.5$^{+0.08}_{-0.07}$	&	0.23$\pm$0.03	&	1.1 (394)		&	1.49$\pm$0.08	&	2.48$\pm$0.02	&	1.12 (394)	&	19.17$\pm$0.21	\\
30352072	&	2008-02-13	&	979.112	&	0.41$^{+0.06}_{-0.07}$	&	0.23$\pm$0.03	&	1.05 (412)		&	1.45$\pm$0.1	&	2.49$\pm$0.01	&	1.07 (412)	&	18.91$\pm$0.2	\\
30352073	&	2008-02-13	&	629.107	&	0.54$^{+0.07}_{-0.09}$	&	0.25$\pm$0.04	&	0.98 (352)		&	1.47$\pm$0.09	&	2.51$\pm$0.02	&	1 (352)	&	19.13$\pm$0.25	\\
30352074-Orb1	&	2008-02-14	&	790.9	&	0.63$^{+0.08}_{-0.11}$	&	0.2$\pm$0.04	&	1.05 (378)		&	1.56$\pm$0.08	&	2.44$\pm$0.03	&	1.06 (378)	&	22.21$\pm$0.28	\\
30352074-Orb2	&	2008-02-14	&	422.7	&	0.74$^{+0.11}_{-0.14}$	&	0.24$\pm$0.06	&	0.93 (273)		&	1.51$\pm$0.1	&	2.49$\pm$0.04	&	0.93 (273)	&	22.38$\pm$0.45	\\
30352075	&	2008-02-14	&	1529.124	&	0.66$^{+0.05}_{-0.06}$	&	0.23$\pm$0.03	&	1.1 (489)		&	1.53$\pm$0.05	&	2.47$\pm$0.02	&	1.11 (489)	&	18.3$\pm$0.15	\\
30352077	&	2008-02-15	&	1589.125	&	0.72$^{+0.07}_{-0.09}$	&	0.24$\pm$0.04	&	0.93 (375)		&	1.51$\pm$0.07	&	2.48$\pm$0.03	&	0.93 (375)	&	17.03$\pm$0.21	\\
30352079	&	2008-02-16	&	1228.244	&	0.47$^{+0.06}_{-0.08}$	&	0.26$\pm$0.04	&	1.12 (365)		&	1.43$\pm$0.1	&	2.52$\pm$0.02	&	1.14 (365)	&	16.74$\pm$0.2	\\
30352081-Orb1	&	2008-03-13	&	461	&	0.49$^{+0.07}_{-0.10}$	&	0.32$\pm$0.06	&	1.05 (276)		&	1.3$\pm$0.14	&	2.59$\pm$0.03	&	1.04 (276)	&	11.75$\pm$0.21	\\
30352083	&	2008-04-02	&	894.619	&	0.54$^{+0.08}_{-0.11}$	&	0.2$\pm$0.04	&	0.97 (382)		&	1.55$\pm$0.09	&	2.45$\pm$0.02	&	0.99 (382)	&	13.39$\pm$0.17	\\
30352085	&	2008-04-03	&	1139.622	&	3.11$^{+0.33}_{-0.26}$	&	0.21$\pm$0.02	&	1.17 (574)		&	1.55$\pm$0.01	&	2.47$\pm$0.05	&	1.17 (574)	&	30$\pm$0.25	\\
30352086	&	2008-04-03	&	1159.607	&	1.35$^{+0.05}_{-0.06}$	&	0.26$\pm$0.03	&	1.02 (511)		&	1.51$\pm$0.03	&	2.49$\pm$0.03	&	1.02 (511)	&	24.79$\pm$0.22	\\
30352087-Orb1	&	2008-04-04	&	360	&	1.67$^{+0.18}_{-0.15}$	&	0.21$\pm$0.04	&	0.97 (358)		&	1.56$\pm$0.05	&	2.44$\pm$0.06	&	0.97 (358)	&	24.45$\pm$0.38	\\
30352087-Orb2	&	2008-04-04	&	758.3	&	1.31$^{+0.08}_{-0.08}$	&	0.21$\pm$0.03	&	1.17 (457)		&	1.57$\pm$0.04	&	2.43$\pm$0.03	&	1.17 (457)	&	23.45$\pm$0.25	\\
30352088	&	2008-04-05	&	1139.21	&	4.1$^{+0.86}_{-0.89}$	&	0.11$\pm$0.03	&	1.07 (503)		&	1.67$\pm$0.02	&	2.36$\pm$0.08	&	0.08 (503)	&	20$\pm$0.21	\\
30352089	&	2008-04-06	&	1019.621	&	1.67$^{+0.09}_{-0.09}$	&	0.24$\pm$0.03	&	1.04 (481)		&	1.53$\pm$0.03	&	2.47$\pm$0.04	&	1.04 (481)	&	20$\pm$0.2	\\
30352091	&	2008-04-08	&	1074.608	&	1.76$^{+0.17}_{-0.14}$	&	0.19$\pm$0.03	&	1.16 (447)		&	1.59$\pm$0.04	&	2.41$\pm$0.05	&	1.16 (447)	&	23.17$\pm$0.28	\\
30352093	&	2008-04-09	&	2197.915	&	1.22$^{+0.05}_{-0.06}$	&	0.2$\pm$0.02	&	1.06 (533)		&	1.58$\pm$0.03	&	2.43$\pm$0.02	&	1.06 (533)	&	17.13$\pm$0.14	\\
30352094-Orb1	&	2008-04-11	&	690.4	&	0.75$^{+0.07}_{-0.09}$	&	0.25$\pm$0.04	&	1.02 (346)		&	1.51$\pm$0.07	&	2.49$\pm$0.03	&	1.03 (346)	&	23.09$\pm$0.32	\\
30352094-Orb2	&	2008-04-11	&	1120	&	0.59$^{+0.13}_{-0.12}$	&	0.18$\pm$0.06	&	0.95 (284)		&	1.55$\pm$0.12	&	2.43$\pm$0.04	&	0.95 (284)	&	11.48$\pm$0.22	\\
30352095	&	2008-04-13	&	2917.377	&	0.7$^{+0.03}_{-0.05}$	&	0.26$\pm$0.02	&	1.12 (514)		&	1.49$\pm$0.04	&	2.5$\pm$0.02	&	1.15 (514)	&	11.56$\pm$0.09	\\
31202001-Orb1	&	2008-05-02	&	901.6	&	0.47$^{+0.08}_{-0.08}$	&	0.19$\pm$0.06	&	1.06 (281)		&	1.5$\pm$0.1	&	2.45$\pm$0.03	&	1.06 (281)	&	6.65$\pm$0.12	\\
30352097	&	2008-05-04	&	734.619	&	1.03$^{+0.08}_{-0.10}$	&	0.21$\pm$0.04	&	1.04 (386)		&	1.56$\pm$0.05	&	2.44$\pm$0.04	&	1.04 (386)	&	16.74$\pm$0.22	\\
30352098-Orb1	&	2008-05-05	&	1146	&	0.54$^{+0.04}_{-0.06}$	&	0.25$\pm$0.03	&	1.16 (461)		&	1.47$\pm$0.06	&	2.51$\pm$0.02	&	1.19 (461)	&	19.42$\pm$0.17	\\
30352098-Orb2	&	2008-05-05	&	846.2	&	0.48$^{+0.05}_{-0.07}$	&	0.27$\pm$0.03	&	1.08 (398)		&	1.42$\pm$0.09	&	2.53$\pm$0.02	&	1.1 (398)	&	17.84$\pm$0.19	\\
31202003	&	2008-05-07	&	1569.056	&	0.41$^{+0.04}_{-0.06}$	&	0.26$\pm$0.03	&	1 (419)		&	1.41$\pm$0.09	&	2.53$\pm$0.01	&	1.04 (419)	&	13.32$\pm$0.12	\\
30352100-Orb1	&	2008-11-08	&	943.8	&	0.5$^{+0.05}_{-0.06}$	&	0.36$\pm$0.04	&	1.16 (330)		&	1.29$\pm$0.1	&	2.63$\pm$0.02	&	1.19 (330)	&	10$\pm$0.13	\\
30352102-Orb1	&	2008-12-03	&	874.8	&	0.75$^{+0.05}_{-0.06}$	&	0.38$\pm$0.04	&	1.12 (339)		&	1.38$\pm$0.07	&	2.62$\pm$0.03	&	1.14 (339)	&	10.71$\pm$0.14	\\
30352102-Orb2	&	2008-12-03	&	1614	&	0.69$^{+0.03}_{-0.04}$	&	0.41$\pm$0.03	&	1.1 (400)		&	1.34$\pm$0.05	&	2.65$\pm$0.02	&	1.15 (400)	&	10$\pm$0.1	\\
30352104	&	2009-01-12	&	794.227	&	1.03$^{+0.06}_{-0.07}$	&	0.34$\pm$0.05	&	1.11 (327)		&	1.44$\pm$0.06	&	2.57$\pm$0.04	&	1.12 (327)	&	10.69$\pm$0.16	\\
30352105	&	2009-01-13	&	769.621	&	0.79$^{+0.08}_{-0.11}$	&	0.25$\pm$0.05	&	1.05 (320)		&	1.51$\pm$0.08	&	2.49$\pm$0.04	&	1.05 (320)	&	10.66$\pm$0.17	\\
30352107	&	2009-01-27	&	929.225	&	0.87$^{+0.08}_{-0.10}$	&	0.24$\pm$0.04	&	1 (349)		&	1.53$\pm$0.07	&	2.47$\pm$0.04	&	1.01 (349)	&	10$\pm$0.15	\\
30352124	&	2009-04-11	&	533.855	&	0.53$^{+0.06}_{-0.08}$	&	0.33$\pm$0.05	&	0.93 (294)		&	1.36$\pm$0.11	&	2.59$\pm$0.03	&	0.94 (294)	&	12.81$\pm$0.2	\\
30352125	&	2009-04-14	&	1019.226	&	0.53$^{+0.06}_{-0.08}$	&	0.26$\pm$0.04	&	1.14 (364)		&	1.44$\pm$0.09	&	2.52$\pm$0.02	&	1.15 (364)	&	11.18$\pm$0.14	\\
30352129	&	2009-04-22	&	724.224	&	0.57$^{+0.08}_{-0.10}$	&	0.28$\pm$0.05	&	1.07 (294)		&	1.43$\pm$0.11	&	2.54$\pm$0.03	&	1.07 (294)	&	9.13$\pm$0.15	\\
30352151	&	2009-05-27	&	974.447	&	0.45$^{+0.06}_{-0.08}$	&	0.3$\pm$0.05	&	1.07 (317)		&	1.29$\pm$0.13	&	2.57$\pm$0.02	&	1.07 (317)	&	8.46$\pm$0.12	\\
31540001-Orb1	&	2009-11-14	&	669.8	&	1.75$^{+0.17}_{-0.14}$	&	0.19$\pm$0.03	&	1.06 (431)		&	1.58$\pm$0.04	&	2.42$\pm$0.05	&	1.06 (431)	&	21.76$\pm$0.27	\\
31540001-Orb3	&	2009-11-14	&	185.7	&	1.61$^{+0.28}_{-0.20}$	&	0.31$\pm$0.1	&	1.09 (191)		&	1.46$\pm$0.09	&	2.54$\pm$0.11	&	1.09 (191)	&	11.36$\pm$0.38	\\
31540002-Orb1	&	2009-11-15	&	602.6	&	2.01$^{+0.27}_{-0.20}$	&	0.2$\pm$0.04	&	0.96 (377)		&	1.56$\pm$0.04	&	2.44$\pm$0.06	&	0.96 (377)	&	17.24$\pm$0.26	\\
31540002-Orb3	&	2009-11-15	&	209.7	&	1.84$^{+0.24}_{-0.22}$	&	0.24$\pm$0.07	&	1.19 (255)		&	1.53$\pm$0.06	&	2.47$\pm$0.09	&	1.19 (255)	&	18.51$\pm$0.45	\\
30352160	&	2009-12-09	&	1024.454	&	0.45$^{+0.06}_{-0.08}$	&	0.3$\pm$0.05	&	1.06 (318)		&	1.33$\pm$0.13	&	2.57$\pm$0.02	&	1.07 (318)	&	8.39$\pm$0.12	\\
30352161	&	2009-12-11	&	969.613	&	0.52$^{+0.06}_{-0.08}$	&	0.29$\pm$0.04	&	1 (328)		&	1.41$\pm$0.1	&	2.55$\pm$0.02	&	1.02 (328)	&	9.41$\pm$0.13	\\
30352162	&	2009-12-13	&	994.602	&	0.59$^{+0.04}_{-0.06}$	&	0.38$\pm$0.04	&	0.95 (332)		&	1.32$\pm$0.08	&	2.64$\pm$0.02	&	0.97 (332)	&	10$\pm$0.12	\\
30352163-Seg1	&	2009-12-15	&	498.6	&	0.8$^{+0.05}_{-0.07}$	&	0.35$\pm$0.05	&	1.1 (316)		&	1.41$\pm$0.07	&	2.59$\pm$0.03	&	1.12 (316)	&	15.28$\pm$0.23	\\
30352163-Seg2	&	2009-12-15	&	490.8	&	0.82$^{+0.05}_{-0.07}$	&	0.37$\pm$0.05	&	1.16 (317)		&	1.4$\pm$0.07	&	2.61$\pm$0.03	&	1.17 (317)	&	15.14$\pm$0.23	\\

\hline
\end{tabular}
\end{center}
\end{table*}


\begin{table*}
\begin{center}
\label{tab:continued}         
\begin{tabular}{l c c c c c c c c c }
\hline
 ObsID	&	Date of Obs.	&	Exposure	&		&	\emph{eplogpar}	&		&		&	\emph{sbpl}	&	 & 	Flux$_{0.3-10.0\,\rm kev}$\\
 
 \cline{4-6}
 \cline{7-9}
 
  		&		&	(sec)	&	$\epsilon_p$\,(keV)	&	$\beta$	&	$\tilde{\chi_{red}^2}$\,(dof)	&	$\Gamma_{\rm low}$	&	$\Gamma_{\rm high}$		&	$\tilde{\chi_{red}^2}$\,(dof) & ( $10^{-10} \rm erg\,cm^{-2} s^{-1}$)	\\

\hline
\hline
30352167	&	2009-12-25	&	1164.605	&	0.76$^{+0.05}_{-0.06}$	&	0.29$\pm$0.03	&	1.02 (420)		&	1.46$\pm$0.05	&	2.53$\pm$0.02	&	1.03 (420)	&	14.16$\pm$0.15	\\
30352168-Seg1	&	2009-12-27	&	597.8	&	0.88$^{+0.08}_{-0.10}$	&	0.26$\pm$0.05	&	1.16 (332)		&	1.51$\pm$0.07	&	2.49$\pm$0.04	&	1.16 (332)	&	13.32$\pm$0.2	\\
30352168-Seg2	&	2009-12-27	&	584.8	&	1.17$^{+0.07}_{-0.08}$	&	0.29$\pm$0.05	&	1.05 (347)		&	1.49$\pm$0.05	&	2.51$\pm$0.04	&	1.06 (347)	&	14.1$\pm$0.22	\\
30352170	&	2009-12-31	&	1174.448	&	1.32$^{+0.05}_{-0.06}$	&	0.24$\pm$0.03	&	1.14 (491)		&	1.53$\pm$0.03	&	2.47$\pm$0.03	&	1.15 (491)	&	21.64$\pm$0.2	\\
30352173-Seg1	&	2010-01-06	&	499.7	&	1.16$^{+0.08}_{-0.09}$	&	0.27$\pm$0.05	&	1.08 (349)		&	1.5$\pm$0.05	&	2.5$\pm$0.04	&	1.08 (349)	&	16.9$\pm$0.26	\\
30352173-Seg2	&	2010-01-06	&	459.1	&	1.33$^{+0.07}_{-0.08}$	&	0.34$\pm$0.05	&	1.18 (338)		&	1.44$\pm$0.05	&	2.57$\pm$0.04	&	1.18 (338)	&	16.81$\pm$0.27	\\
30352178-Seg1	&	2010-01-14	&	699.6	&	1.15$^{+0.06}_{-0.06}$	&	0.31$\pm$0.03	&	1.17 (419)		&	1.47$\pm$0.04	&	2.54$\pm$0.03	&	1.17 (419)	&	20.96$\pm$0.24	\\
30352178-Seg2	&	2010-01-14	&	592.4	&	1.15$^{+0.07}_{-0.07}$	&	0.28$\pm$0.04	&	1.01 (400)		&	1.49$\pm$0.04	&	2.51$\pm$0.03	&	1.02 (400)	&	21.17$\pm$0.27	\\
30352179-Seg1	&	2010-01-15	&	498.6	&	1.63$^{+0.08}_{-0.08}$	&	0.33$\pm$0.04	&	1.12 (402)		&	1.44$\pm$0.03	&	2.56$\pm$0.04	&	1.12 (402)	&	22.89$\pm$0.3	\\
30352179-Seg2	&	2010-01-15	&	497	&	1.72$^{+0.09}_{-0.08}$	&	0.35$\pm$0.04	&	1.17 (402)		&	1.42$\pm$0.03	&	2.58$\pm$0.04	&	1.17 (402)	&	23.03$\pm$0.3	\\
30352182	&	2010-01-16	&	939.454	&	1.27$^{+0.14}_{-0.15}$	&	0.21$\pm$0.06	&	1.06 (300)		&	1.56$\pm$0.07	&	2.44$\pm$0.06	&	1.06 (300)	&	15.53$\pm$0.31	\\
30352187	&	2010-01-20	&	704.617	&	1.53$^{+0.15}_{-0.13}$	&	0.18$\pm$0.04	&	1.05 (402)		&	1.6$\pm$0.04	&	2.4$\pm$0.05	&	1.05 (402)	&	16.87$\pm$0.23	\\
31540008	&	2010-01-21	&	1059.621	&	0.68$^{+0.09}_{-0.12}$	&	0.2$\pm$0.04	&	1.09 (362)		&	1.56$\pm$0.08	&	2.44$\pm$0.03	&	1.1 (362)	&	10.55$\pm$0.14	\\
31540010	&	2010-01-23	&	1109.454	&	0.72$^{+0.08}_{-0.10}$	&	0.25$\pm$0.05	&	0.93 (335)		&	1.51$\pm$0.08	&	2.49$\pm$0.03	&	0.94 (335)	&	7.62$\pm$0.11	\\
30352188	&	2010-01-24	&	1124.621	&	0.81$^{+0.21}_{-0.22}$	&	0.11$\pm$0.05	&	0.95 (308)		&	1.69$\pm$0.12	&	2.31$\pm$0.06	&	0.95 (308)	&	12.25$\pm$0.24	\\
30352189	&	2010-01-27	&	1104.609	&	0.83$^{+0.05}_{-0.07}$	&	0.33$\pm$0.04	&	1.08 (360)		&	1.42$\pm$0.06	&	2.57$\pm$0.03	&	1.08 (360)	&	9.42$\pm$0.13	\\
30352196-Seg1	&	2010-02-05	&	598.2	&	1.63$^{+0.15}_{-0.16}$	&	0.19$\pm$0.04	&	1.12 (354)		&	1.58$\pm$0.05	&	2.42$\pm$0.06	&	1.12 (354)	&	33.36$\pm$0.53	\\
30352196-Seg2	&	2010-02-05	&	585.2	&	1.6$^{+0.16}_{-0.14}$	&	0.22$\pm$0.05	&	1.18 (345)		&	1.55$\pm$0.05	&	2.45$\pm$0.06	&	1.19 (345)	&	33.15$\pm$0.54	\\
30352200	&	2010-02-10	&	1014.608	&	1.07$^{+0.06}_{-0.08}$	&	0.29$\pm$0.04	&	0.99 (358)		&	1.48$\pm$0.05	&	2.52$\pm$0.04	&	0.99 (358)	&	10$\pm$0.14	\\
30352203-Seg1	&	2010-02-12	&	596.2	&	1.2$^{+0.06}_{-0.07}$	&	0.28$\pm$0.04	&	1.05 (390)		&	1.49$\pm$0.04	&	2.52$\pm$0.04	&	1.05 (390)	&	20$\pm$0.26	\\
30352203-Seg2	&	2010-02-12	&	510.6	&	1.25$^{+0.06}_{-0.07}$	&	0.3$\pm$0.04	&	1.18 (373)		&	1.47$\pm$0.05	&	2.53$\pm$0.04	&	1.18 (373)	&	20$\pm$0.28	\\
30352205	&	2010-02-15	&	929.527	&	0.85$^{+0.06}_{-0.08}$	&	0.23$\pm$0.03	&	0.99 (412)		&	1.53$\pm$0.05	&	2.47$\pm$0.03	&	1 (412)	&	18.57$\pm$0.21	\\
30352206-Seg1	&	2010-02-16	&	598.1	&	3.46$^{+0.38}_{-0.32}$	&	0.21$\pm$0.03	&	1.18 (514)		&	1.54$\pm$0.02	&	2.49$\pm$0.06	&	1.17 (514)	&	35.7$\pm$0.38	\\
30352206-Seg2	&	2010-02-16	&	586.9	&	3.38$^{+0.36}_{-0.31}$	&	0.22$\pm$0.03	&	1.09 (505)		&	1.53$\pm$0.02	&	2.49$\pm$0.06	&	1.09 (505)	&	34.99$\pm$0.38	\\
30352207	&	2010-02-17	&	1064.609	&	6.58$^{+1.41}_{-1.44}$	&	0.17$\pm$0.03	&	0.99 (515)		&	1.58$\pm$0.01	&	2.51$\pm$0.14	&	1 (515)	&	46.21$\pm$0.56	\\
30352208	&	2010-02-18	&	607.231	&	2.42$^{+0.48}_{-0.30}$	&	0.19$\pm$0.04	&	1.09 (384)		&	1.57$\pm$0.04	&	2.43$\pm$0.07	&	1.09 (384)	&	31.27$\pm$0.49	\\
30352212	&	2010-02-24	&	1419.578	&	3.1$^{+0.30}_{-0.39}$	&	0.18$\pm$0.03	&	1.07 (496)		&	1.58$\pm$0.02	&	2.43$\pm$0.06	&	1.07 (496)	&	17.93$\pm$0.2	\\
30352213	&	2010-02-25	&	884.617	&	4.07$^{+0.78}_{-0.77}$	&	0.14$\pm$0.03	&	1.07 (485)		&	1.63$\pm$0.02	&	2.4$\pm$0.08	&	1.06 (485)	&	20$\pm$0.23	\\
30352214	&	2010-02-26	&	1389.166	&	4.43$^{+0.64}_{-0.47}$	&	0.22$\pm$0.02	&	1.14 (578)		&	1.52$\pm$0.01	&	2.57$\pm$0.07	&	1.11 (578)	&	24.95$\pm$0.22	\\
30352215	&	2010-02-27	&	1278.909	&	4.06$^{+0.73}_{-0.69}$	&	0.15$\pm$0.03	&	0.93 (496)		&	1.61$\pm$0.02	&	2.43$\pm$0.08	&	0.92 (496)	&	18.46$\pm$0.21	\\
30352217-Seg1	&	2010-03-02	&	498.9	&	5.94$^{+1.20}_{-0.80}$	&	0.17$\pm$0.04	&	1.09 (436)		&	1.58$\pm$0.02	&	2.55$\pm$0.17	&	1.08 (436)	&	25.35$\pm$0.37	\\
30352219	&	2010-03-04	&	899.62	&	2.25$^{+0.36}_{-0.24}$	&	0.19$\pm$0.04	&	1.07 (417)		&	1.57$\pm$0.03	&	2.43$\pm$0.06	&	1.07 (417)	&	24.84$\pm$0.35	\\
30352221	&	2010-03-06	&	1194.607	&	3.01$^{+0.88}_{-0.52}$	&	0.13$\pm$0.03	&	1.07 (452)		&	1.65$\pm$0.03	&	2.35$\pm$0.07	&	1.07 (452)	&	14.65$\pm$0.18	\\
30352222	&	2010-03-07	&	1189.619	&	4.37$^{+1.02}_{-0.98}$	&	0.16$\pm$0.04	&	1.06 (390)		&	1.6$\pm$0.03	&	2.44$\pm$0.13	&	1.06 (390)	&	14.78$\pm$0.25	\\
30352224	&	2010-03-09	&	969.617	&	4.4$^{+0.76}_{-0.53}$	&	0.23$\pm$0.03	&	1.07 (520)		&	1.5$\pm$0.01	&	2.6$\pm$0.09	&	1.05 (520)	&	23.3$\pm$0.25	\\
30352226	&	2010-03-11	&	944.696	&	1.5$^{+0.09}_{-0.09}$	&	0.24$\pm$0.03	&	1.06 (433)		&	1.53$\pm$0.04	&	2.47$\pm$0.04	&	1.07 (433)	&	16.25$\pm$0.19	\\
30352227	&	2010-03-12	&	1014.446	&	1.72$^{+0.09}_{-0.09}$	&	0.24$\pm$0.03	&	1 (491)		&	1.53$\pm$0.03	&	2.47$\pm$0.04	&	1 (491)	&	22.23$\pm$0.22	\\
30352228	&	2010-03-14	&	959.448	&	0.45$^{+0.07}_{-0.09}$	&	0.26$\pm$0.04	&	0.96 (326)		&	1.42$\pm$0.12	&	2.53$\pm$0.02	&	0.98 (326)	&	10$\pm$0.14	\\
30352233	&	2010-03-20	&	1344.621	&	0.73$^{+0.07}_{-0.09}$	&	0.22$\pm$0.04	&	0.94 (387)		&	1.54$\pm$0.07	&	2.46$\pm$0.03	&	0.95 (387)	&	9.38$\pm$0.11	\\
30352234	&	2010-03-21	&	1039.564	&	0.62$^{+0.12}_{-0.16}$	&	0.15$\pm$0.04	&	1.01 (366)		&	1.62$\pm$0.09	&	2.38$\pm$0.03	&	1.01 (366)	&	10$\pm$0.14	\\
30352235	&	2010-03-22	&	1009.448	&	0.66$^{+0.06}_{-0.07}$	&	0.33$\pm$0.05	&	0.99 (328)		&	1.4$\pm$0.08	&	2.58$\pm$0.03	&	1 (328)	&	8.76$\pm$0.12	\\
30352236	&	2010-03-23	&	979.606	&	0.66$^{+0.15}_{-0.21}$	&	0.17$\pm$0.06	&	0.78 (283)		&	1.6$\pm$0.12	&	2.4$\pm$0.04	&	0.78 (283)	&	8.55$\pm$0.17	\\
30352237-Seg1	&	2010-03-25	&	497.5	&	1.28$^{+0.16}_{-0.17}$	&	0.18$\pm$0.06	&	0.99 (298)		&	1.59$\pm$0.07	&	2.41$\pm$0.06	&	0.99 (298)	&	12.23$\pm$0.24	\\
30352237-Seg2	&	2010-03-25	&	592.6	&	1.3$^{+0.13}_{-0.14}$	&	0.2$\pm$0.05	&	1.09 (324)		&	1.57$\pm$0.06	&	2.43$\pm$0.06	&	1.09 (324)	&	12.48$\pm$0.22	\\
30352238	&	2010-03-27	&	1049.452	&	1.7$^{+0.10}_{-0.10}$	&	0.27$\pm$0.04	&	0.97 (431)		&	1.49$\pm$0.03	&	2.51$\pm$0.04	&	0.98 (431)	&	14.52$\pm$0.18	\\
30352239	&	2010-03-29	&	1009.634	&	2.42$^{+0.35}_{-0.24}$	&	0.24$\pm$0.04	&	0.95 (395)		&	1.52$\pm$0.03	&	2.48$\pm$0.07	&	0.95 (395)	&	15.11$\pm$0.23	\\
30352240	&	2010-03-31	&	1024.625	&	1.6$^{+0.11}_{-0.11}$	&	0.24$\pm$0.04	&	1.06 (413)		&	1.53$\pm$0.04	&	2.47$\pm$0.05	&	1.06 (413)	&	12.42$\pm$0.16	\\
30352241	&	2010-03-31	&	691.235	&	1.48$^{+0.14}_{-0.13}$	&	0.25$\pm$0.05	&	1.05 (317)		&	1.52$\pm$0.05	&	2.48$\pm$0.06	&	1.05 (317)	&	10.5$\pm$0.19	\\
30352242	&	2010-04-01	&	849.555	&	0.9$^{+0.09}_{-0.11}$	&	0.24$\pm$0.05	&	0.92 (317)		&	1.53$\pm$0.07	&	2.48$\pm$0.04	&	0.93 (317)	&	8.84$\pm$0.14	\\
31202006	&	2010-04-02	&	1069.602	&	0.97$^{+0.08}_{-0.10}$	&	0.21$\pm$0.04	&	1 (379)		&	1.56$\pm$0.06	&	2.45$\pm$0.04	&	1.01 (379)	&	10$\pm$0.14	\\
31202007	&	2010-04-03	&	869.227	&	1.65$^{+0.14}_{-0.12}$	&	0.27$\pm$0.05	&	1.08 (342)		&	1.5$\pm$0.05	&	2.5$\pm$0.06	&	1.08 (342)	&	11.76$\pm$0.2	\\
31202008	&	2010-04-05	&	808.992	&	0.66$^{+0.21}_{-0.35}$	&	0.13$\pm$0.07	&	1.02 (262)		&	1.63$\pm$0.16	&	2.36$\pm$0.06	&	1.02 (262)	&	7.77$\pm$0.18	\\
31202009-Seg1	&	2010-04-07	&	581.4	&	0.71$^{+0.19}_{-0.30}$	&	0.23$\pm$0.11	&	1.1 (183)		&	1.51$\pm$0.19	&	2.47$\pm$0.07	&	1.1 (183)	&	7.77$\pm$0.27	\\
31202009-Seg2	&	2010-04-07	&	519.9	&	0.92$^{+0.20}_{-0.29}$	&	0.24$\pm$0.11	&	1.11 (171)		&	1.52$\pm$0.17	&	2.48$\pm$0.1	&	1.11 (171)	&	8.12$\pm$0.31	\\
31202015	&	2010-04-23	&	789.448	&	0.62$^{+0.08}_{-0.11}$	&	0.24$\pm$0.05	&	0.99 (317)		&	1.49$\pm$0.09	&	2.49$\pm$0.03	&	0.99 (317)	&	10$\pm$0.15	\\
31202017	&	2010-04-28	&	884.609	&	1.38$^{+0.08}_{-0.09}$	&	0.25$\pm$0.04	&	1.16 (409)		&	1.52$\pm$0.04	&	2.48$\pm$0.04	&	1.17 (409)	&	14.05$\pm$0.18	\\
31202018	&	2010-04-29	&	994.752	&	1.71$^{+0.11}_{-0.10}$	&	0.25$\pm$0.03	&	1.12 (444)		&	1.52$\pm$0.03	&	2.48$\pm$0.04	&	1.12 (444)	&	15.18$\pm$0.18	\\
31202019-Seg1	&	2010-05-02	&	598.3	&	1.27$^{+0.12}_{-0.12}$	&	0.21$\pm$0.05	&	1 (334)		&	1.56$\pm$0.05	&	2.44$\pm$0.05	&	1 (334)	&	13.36$\pm$0.22	\\
31202019-Seg2	&	2010-05-02	&	589	&	1.46$^{+0.11}_{-0.13}$	&	0.25$\pm$0.05	&	0.99 (339)		&	1.52$\pm$0.05	&	2.48$\pm$0.05	&	0.99 (339)	&	13.64$\pm$0.23	\\
31202020	&	2010-05-03	&	1194.428	&	0.59$^{+0.06}_{-0.08}$	&	0.24$\pm$0.04	&	1.11 (391)		&	1.5$\pm$0.07	&	2.49$\pm$0.02	&	1.12 (391)	&	12.24$\pm$0.14	\\

\hline
\end{tabular}
\end{center}
\end{table*}


\begin{table*}
\begin{center}
\label{tab:continued}         
\begin{tabular}{l c c c c c c c c c }
\hline
 ObsID	&	Date of Obs.	&	Exposure	&		&	\emph{eplogpar}	&		&		&	\emph{sbpl}	&	 & 	Flux$_{0.3-10.0\,\rm kev}$\\
 
 \cline{4-6}
 \cline{7-9}
 
  		&		&	(sec)	&	$\epsilon_p$\,(keV)	&	$\beta$	&	$\tilde{\chi_{red}^2}$\,(dof)	&	$\Gamma_{\rm low}$	&	$\Gamma_{\rm high}$		&	$\tilde{\chi_{red}^2}$\,(dof) & ( $10^{-10} \rm erg\,cm^{-2} s^{-1}$)	\\

\hline
\hline
31202021-Seg1	&	2010-05-04	&	399.7	&	1.26$^{+0.09}_{-0.09}$	&	0.31$\pm$0.05	&	1.08 (304)		&	1.47$\pm$0.06	&	2.54$\pm$0.05	&	1.08 (304)	&	15.03$\pm$0.27	\\
31202021-Seg2	&	2010-05-04	&	399.7	&	1.23$^{+0.10}_{-0.10}$	&	0.27$\pm$0.05	&	0.98 (304)		&	1.5$\pm$0.06	&	2.5$\pm$0.05	&	0.99 (304)	&	15.18$\pm$0.28	\\
31202021-Seg3	&	2010-05-04	&	373.9	&	1.2$^{+0.09}_{-0.10}$	&	0.29$\pm$0.06	&	0.88 (295)		&	1.48$\pm$0.06	&	2.52$\pm$0.05	&	0.88 (295)	&	15.13$\pm$0.28	\\
31202022	&	2010-05-05	&	1194.454	&	1.89$^{+0.14}_{-0.12}$	&	0.23$\pm$0.03	&	1.01 (469)		&	1.54$\pm$0.03	&	2.46$\pm$0.04	&	1.01 (469)	&	15.49$\pm$0.17	\\
31202023	&	2010-05-06	&	1014.46	&	3.23$^{+0.78}_{-0.47}$	&	0.16$\pm$0.03	&	1.11 (468)		&	1.6$\pm$0.02	&	2.41$\pm$0.07	&	1.11 (468)	&	16.5$\pm$0.2	\\
31202026	&	2010-05-09	&	1149.455	&	0.68$^{+0.03}_{-0.04}$	&	0.42$\pm$0.03	&	0.99 (410)		&	1.32$\pm$0.05	&	2.67$\pm$0.02	&	1.03 (410)	&	14.41$\pm$0.14	\\
31202027	&	2010-05-10	&	954.443	&	0.44$^{+0.05}_{-0.06}$	&	0.35$\pm$0.05	&	1.03 (316)		&	1.25$\pm$0.13	&	2.62$\pm$0.02	&	1.06 (316)	&	9.37$\pm$0.12	\\
31202028	&	2010-05-11	&	994.607	&	0.87$^{+0.04}_{-0.06}$	&	0.31$\pm$0.04	&	1.07 (409)		&	1.45$\pm$0.05	&	2.55$\pm$0.03	&	1.08 (409)	&	14.07$\pm$0.16	\\
31202029	&	2010-05-14	&	854.617	&	0.44$^{+0.08}_{-0.10}$	&	0.28$\pm$0.05	&	1.01 (292)		&	1.36$\pm$0.15	&	2.54$\pm$0.02	&	1.02 (292)	&	7.52$\pm$0.12	\\
31202031	&	2010-05-20	&	1124.602	&	0.73$^{+0.07}_{-0.09}$	&	0.25$\pm$0.04	&	0.99 (348)		&	1.49$\pm$0.07	&	2.5$\pm$0.03	&	0.99 (348)	&	8.46$\pm$0.12	\\
31202033	&	2010-05-26	&	1009.618	&	0.53$^{+0.12}_{-0.15}$	&	0.17$\pm$0.04	&	0.94 (340)		&	1.58$\pm$0.11	&	2.4$\pm$0.03	&	0.95 (340)	&	10$\pm$0.15	\\
31202034	&	2010-05-29	&	1164.461	&	0.42$^{+0.05}_{-0.06}$	&	0.33$\pm$0.04	&	0.96 (330)		&	1.21$\pm$0.14	&	2.61$\pm$0.02	&	0.96 (330)	&	8.56$\pm$0.11	\\
31202050	&	2010-12-13	&	999.639	&	0.49$^{+0.04}_{-0.06}$	&	0.45$\pm$0.05	&	1.15 (291)		&	1.19$\pm$0.12	&	2.71$\pm$0.02	&	1.18 (291)	&	7.55$\pm$0.11	\\
31202140	&	2012-05-31	&	1069.606	&	0.43$^{+0.05}_{-0.06}$	&	0.29$\pm$0.04	&	1.17 (377)		&	1.37$\pm$0.1	&	2.56$\pm$0.01	&	1.18 (377)	&	12.85$\pm$0.14	\\
35014054	&	2013-03-23	&	974.451	&	0.55$^{+0.08}_{-0.11}$	&	0.16$\pm$0.03	&	1.01 (439)		&	1.6$\pm$0.07	&	2.39$\pm$0.02	&	1.02 (439)	&	18.46$\pm$0.19	\\
35014056	&	2013-03-30	&	1074.607	&	1.07$^{+0.07}_{-0.08}$	&	0.24$\pm$0.04	&	1.06 (407)		&	1.53$\pm$0.05	&	2.47$\pm$0.04	&	1.07 (407)	&	25.92$\pm$0.33	\\
80050019-Orb1	&	2013-04-12	&	436.9	&	1.88$^{+0.1}_{-0.09}$	&	0.4$\pm$0.04	&	1.24 (419)		&	1.38$\pm$0.03	&	2.62$\pm$0.05	&	1.24 (419)	&	58.7$\pm$0.77	\\
80050019-Orb5	&	2013-04-12	&	1516	&	1.6$^{+0.05}_{-0.05}$	&	0.35$\pm$0.02	&	1.22 (554)		&	1.43$\pm$0.02	&	2.57$\pm$0.02	&	1.23 (554)	&	53.7$\pm$0.41	\\
35014065-Orb3	&	2013-04-17	&	1497	&	1.86$^{+0.17}_{-0.13}$	&	0.14$\pm$0.02	&	1.12 (561)		&	1.65$\pm$0.02	&	2.35$\pm$0.04	&	1.12 (561)	&	25.43$\pm$0.20	\\
35014079	&	2013-12-05	&	1034.449	&	0.45$^{+0.06}_{-0.08}$	&	0.27$\pm$0.04	&	1.17 (352)		&	1.43$\pm$0.1	&	2.55$\pm$0.02	&	1.18 (352)	&	10.67$\pm$0.13	\\
35014081-Seg1	&	2013-12-09	&	498.6	&	1.04$^{+0.08}_{-0.10}$	&	0.25$\pm$0.04	&	1.11 (343)		&	1.53$\pm$0.06	&	2.48$\pm$0.04	&	1.12 (343)	&	17.07$\pm$0.26	\\
35014082	&	2013-12-11	&	1054.607	&	0.76$^{+0.06}_{-0.08}$	&	0.22$\pm$0.03	&	1.11 (430)		&	1.54$\pm$0.06	&	2.46$\pm$0.03	&	1.11 (430)	&	15.62$\pm$0.17	\\
35014083-Seg1	&	2013-12-13	&	408.5	&	2.13$^{+0.54}_{-0.30}$	&	0.15$\pm$0.04	&	1.03 (358)		&	1.63$\pm$0.05	&	2.37$\pm$0.07	&	1.03 (358)	&	21.05$\pm$0.34	\\
35014083-Seg2	&	2013-12-13	&	459	&	1.91$^{+0.22}_{-0.17}$	&	0.23$\pm$0.04	&	1.17 (354)		&	1.53$\pm$0.04	&	2.47$\pm$0.06	&	1.17 (354)	&	19.42$\pm$0.31	\\
35014106	&	2014-02-14	&	1029.617	&	0.77$^{+0.04}_{-0.06}$	&	0.27$\pm$0.03	&	1.12 (459)		&	1.5$\pm$0.05	&	2.51$\pm$0.02	&	1.15 (459)	&	20$\pm$0.19	\\
35014107	&	2014-02-15	&	554.444	&	0.43$^{+0.08}_{-0.10}$	&	0.22$\pm$0.04	&	1.05 (335)		&	1.46$\pm$0.13	&	2.48$\pm$0.02	&	1.06 (335)	&	29.23$\pm$0.41	\\
35014125	&	2014-04-01	&	324.617	&	0.48$^{+0.12}_{-0.16}$	&	0.26$\pm$0.08	&	0.99 (222)		&	1.44$\pm$0.19	&	2.52$\pm$0.04	&	1 (222)	&	9.47$\pm$0.23	\\
35014154-Seg1	&	2015-01-02	&	499.2	&	0.77$^{+0.10}_{-0.13}$	&	0.21$\pm$0.04	&	0.99 (334)		&	1.56$\pm$0.08	&	2.44$\pm$0.04	&	0.99 (334)	&	17.11$\pm$0.26	\\
35014154-Seg2	&	2015-01-02	&	473.3	&	0.97$^{+0.10}_{-0.12}$	&	0.21$\pm$0.04	&	0.94 (340)		&	1.57$\pm$0.07	&	2.43$\pm$0.04	&	0.95 (340)	&	17.59$\pm$0.27	\\
35014157-Orb1	&	2015-01-06	&	564.8	&	0.76$^{+0.07}_{-0.09}$	&	0.29$\pm$0.05	&	1.19 (321)		&	1.46$\pm$0.08	&	2.54$\pm$0.03	&	1.19 (321)	&	14.01$\pm$0.21	\\
35014157-Orb2	&	2015-01-06	&	445.5	&	1.12$^{+0.09}_{-0.10}$	&	0.25$\pm$0.05	&	1.19 (316)		&	1.52$\pm$0.06	&	2.48$\pm$0.05	&	1.18 (316)	&	15.63$\pm$0.26	\\
35014160-Seg1	&	2015-01-13	&	395.6	&	1.31$^{+0.22}_{-0.21}$	&	0.14$\pm$0.05	&	1.03 (313)		&	1.65$\pm$0.08	&	2.35$\pm$0.07	&	1.03 (313)	&	16.16$\pm$0.3	\\
35014160-Seg2	&	2015-01-13	&	554.4	&	1.22$^{+0.13}_{-0.14}$	&	0.17$\pm$0.04	&	1.15 (359)		&	1.61$\pm$0.06	&	2.39$\pm$0.05	&	1.15 (359)	&	16.57$\pm$0.25	\\
35014161-Seg1	&	2015-01-15	&	398	&	0.76$^{+0.11}_{-0.09}$	&	0.31$\pm$0.06	&	0.99 (283)		&	1.44$\pm$0.09	&	2.55$\pm$0.04	&	0.99 (283)	&	13.5$\pm$0.25	\\
35014161-Seg2	&	2015-01-15	&	460.1	&	0.78$^{+0.09}_{-0.08}$	&	0.33$\pm$0.05	&	1.09 (298)		&	1.42$\pm$0.08	&	2.57$\pm$0.04	&	1.09 (298)	&	13.65$\pm$0.23	\\
35014162-Seg1	&	2015-01-17	&	400.7	&	0.72$^{+0.13}_{-0.12}$	&	0.09$\pm$0.05	&	1.1 (301)		&	1.71$\pm$0.09	&	2.29$\pm$0.06	&	1.11 (301)	&	14.39$\pm$0.27	\\
35014162-Seg2	&	2015-01-17	&	301.3	&	0.77$^{+0.13}_{-0.12}$	&	0.17$\pm$0.06	&	1.11 (265)		&	1.6$\pm$0.08	&	2.4$\pm$0.05	&	1.11 (265)	&	14.57$\pm$0.31	\\
35014162-Seg3	&	2015-01-17	&	262.1	&	1.06$^{+0.51}_{-0.32}$	&	0.11$\pm$0.07	&	1.1 (257)		&	1.7$\pm$0.08	&	2.3$\pm$0.08	&	1.11 (257)	&	14.94$\pm$0.36	\\
35014163	&	2015-01-18	&	969.448	&	0.54$^{+0.06}_{-0.08}$	&	0.29$\pm$0.04	&	1.13 (346)		&	1.42$\pm$0.09	&	2.55$\pm$0.02	&	1.15 (346)	&	11.35$\pm$0.15	\\
35014164-Seg1	&	2015-01-19	&	500.3	&	0.93$^{+0.07}_{-0.06}$	&	0.34$\pm$0.04	&	0.93 (347)		&	1.43$\pm$0.06	&	2.58$\pm$0.03	&	0.94 (347)	&	18.5$\pm$0.26	\\
35014164-Seg2	&	2015-01-19	&	475.8	&	1.05$^{+0.06}_{-0.05}$	&	0.42$\pm$0.05	&	1.14 (343)		&	1.37$\pm$0.05	&	2.64$\pm$0.03	&	1.15 (343)	&	18.14$\pm$0.26	\\
35014165-Orb1	&	2015-01-21	&	742.9	&	1.21$^{+0.13}_{-0.14}$	&	0.22$\pm$0.06	&	0.97 (288)		&	1.55$\pm$0.08	&	2.45$\pm$0.06	&	0.98 (288)	&	10$\pm$0.21	\\
35014166-Seg1	&	2015-01-21	&	500.9	&	1.23$^{+0.15}_{-0.16}$	&	0.15$\pm$0.05	&	1.19 (345)		&	1.63$\pm$0.07	&	2.37$\pm$0.05	&	1.2 (345)	&	17.51$\pm$0.28	\\
35014166-Seg2	&	2015-01-21	&	498.7	&	0.97$^{+0.15}_{-0.20}$	&	0.14$\pm$0.04	&	1.18 (344)		&	1.66$\pm$0.08	&	2.35$\pm$0.05	&	1.19 (344)	&	17.6$\pm$0.28	\\
35014166-Seg3	&	2015-01-21	&	462.1	&	1.34$^{+0.18}_{-0.17}$	&	0.16$\pm$0.05	&	1.02 (335)		&	1.62$\pm$0.07	&	2.38$\pm$0.06	&	1.02 (335)	&	17.77$\pm$0.3	\\
35014167	&	2015-01-24	&	329.457	&	0.83$^{+0.17}_{-0.25}$	&	0.18$\pm$0.07	&	0.92 (252)		&	1.59$\pm$0.13	&	2.41$\pm$0.06	&	0.92 (252)	&	13.11$\pm$0.31	\\
35014168	&	2015-01-25	&	1134.604	&	0.47$^{+0.08}_{-0.11}$	&	0.21$\pm$0.04	&	1.1 (361)		&	1.51$\pm$0.11	&	2.46$\pm$0.02	&	1.11 (361)	&	10.52$\pm$0.14	\\
35014170	&	2015-01-29	&	1004.619	&	0.56$^{+0.05}_{-0.07}$	&	0.33$\pm$0.04	&	1.13 (333)		&	1.39$\pm$0.09	&	2.59$\pm$0.02	&	1.16 (333)	&	9.45$\pm$0.13	\\
35014173	&	2015-02-06	&	874.619	&	1.62$^{+0.21}_{-0.17}$	&	0.24$\pm$0.06	&	1.06 (290)		&	1.53$\pm$0.06	&	2.47$\pm$0.08	&	1.06 (290)	&	10.66$\pm$0.23	\\
35014174	&	2015-02-09	&	824.621	&	0.67$^{+0.05}_{-0.07}$	&	0.29$\pm$0.04	&	1.08 (374)		&	1.45$\pm$0.07	&	2.54$\pm$0.02	&	1.1 (374)	&	14.31$\pm$0.17	\\
35014175	&	2015-02-11	&	914.109	&	6.06$^{+1.50}_{-1.80}$	&	0.13$\pm$0.03	&	1.05 (475)		&	1.64$\pm$0.02	&	2.45$\pm$0.13	&	0.13 (475)	&	18.46$\pm$0.22	\\
35014176	&	2015-02-13	&	994.609	&	2.02$^{+0.49}_{-0.28}$	&	0.12$\pm$0.04	&	1.08 (418)		&	1.67$\pm$0.05	&	2.33$\pm$0.06	&	1.08 (418)	&	18.65$\pm$0.25	\\
35014177	&	2015-02-15	&	709.448	&	0.91$^{+0.09}_{-0.11}$	&	0.21$\pm$0.04	&	1.02 (356)		&	1.56$\pm$0.07	&	2.45$\pm$0.04	&	1.02 (356)	&	13.07$\pm$0.19	\\
35014178	&	2015-02-17	&	189.635	&	0.66$^{+0.08}_{-0.08}$	&	0.27$\pm$0.07	&	0.87 (191)		&	1.45$\pm$0.11	&	2.53$\pm$0.06	&	0.87 (191)	&	11$\pm$0.34	\\
35014185	&	2015-03-08	&	844.63	&	1.20$^{+0.05}_{-0.06}$	&	0.29$\pm$0.03	&	0.95 (426)		&	1.48$\pm$0.04	&	2.52$\pm$0.03	&	0.96 (426)	&	20$\pm$0.23	\\
35014186	&	2015-03-11	&	835.035	&	0.54$^{+0.07}_{-0.10}$	&	0.23$\pm$0.04	&	1.05 (367)		&	1.48$\pm$0.09	&	2.49$\pm$0.02	&	1.05 (367)	&	15.18$\pm$0.19	\\
35014188-Seg1	&	2015-03-15	&	497.4	&	0.82$^{+0.12}_{-0.10}$	&	0.21$\pm$0.04	&	1.1 (342)		&	1.56$\pm$0.07	&	2.44$\pm$0.04	&	1.1 (342)	&	17.05$\pm$0.26	\\
35014188-Seg2	&	2015-03-15	&	494.3	&	0.8$^{+0.11}_{-0.10}$	&	0.22$\pm$0.05	&	1.19 (338)		&	1.54$\pm$0.07	&	2.46$\pm$0.04	&	1.19 (338)	&	16.92$\pm$0.26	\\
35014190	&	2015-03-19	&	949.46	&	0.55$^{+0.05}_{-0.06}$	&	0.37$\pm$0.05	&	0.99 (318)		&	1.28$\pm$0.1	&	2.64$\pm$0.02	&	1 (318)	&	10$\pm$0.13	\\
35014194	&	2015-03-27	&	1004.615	&	0.87$^{+0.04}_{-0.06}$	&	0.28$\pm$0.03	&	1.02 (458)		&	1.48$\pm$0.04	&	2.52$\pm$0.02	&	1.04 (458)	&	20$\pm$0.19	\\

\hline
\end{tabular}
\end{center}
\end{table*}

\begin{table*}
\begin{center}
\label{tab:continued}         
\begin{tabular}{l c c c c c c c c c }
\hline
 ObsID	&	Date of Obs.	&	Exposure	&		&	\emph{eplogpar}	&		&		&	\emph{sbpl}	&	 & 	Flux$_{0.3-10.0\,\rm kev}$\\
 
 \cline{4-6}
 \cline{7-9}
 
  &		&	(sec)	&	$\epsilon_p$\,(keV)	&	$\beta$	&	$\tilde{\chi_{red}^2}$\,(dof)	&	$\Gamma_{\rm low}$	&	$\Gamma_{\rm high}$		&	$\tilde{\chi_{red}^2}$\,(dof) & ( $10^{-10} \rm erg\,cm^{-2} s^{-1}$)	\\

\hline
\hline
35014199	&	2015-04-03	&	1119.62	&	0.57$^{+0.05}_{-0.07}$	&	0.32$\pm$0.04	&	1.03 (360)		&	1.4$\pm$0.08	&	2.58$\pm$0.02	&	1.06 (360)	&	13.4$\pm$0.16	\\
35014201-Seg1	&	2015-04-07	&	399.4	&	0.41$^{+0.08}_{-0.10}$	&	0.23$\pm$0.07	&	0.99 (240)		&	1.36$\pm$0.14	&	2.5$\pm$0.03	&	0.98 (240)	&	10$\pm$0.22	\\
35014201-Seg2	&	2015-04-07	&	409.6	&	0.46$^{+0.08}_{-0.10}$	&	0.26$\pm$0.07	&	0.99 (242)		&	1.37$\pm$0.14	&	2.53$\pm$0.03	&	0.99 (242)	&	10$\pm$0.21	\\
35014202	&	2015-04-09	&	919.066	&	0.89$^{+0.05}_{-0.07}$	&	0.28$\pm$0.04	&	1.17 (399)		&	1.49$\pm$0.05	&	2.51$\pm$0.03	&	1.18 (399)	&	15.9$\pm$0.19	\\
35014203-Seg1	&	2015-04-11	&	600.2	&	1.21$^{+0.06}_{-0.08}$	&	0.29$\pm$0.04	&	1.05 (376)		&	1.49$\pm$0.05	&	2.52$\pm$0.04	&	1.06 (376)	&	18.44$\pm$0.25	\\
35014203-Seg2	&	2015-04-11	&	527.4	&	1.28$^{+0.08}_{-0.09}$	&	0.27$\pm$0.04	&	1.11 (360)		&	1.51$\pm$0.05	&	2.5$\pm$0.04	&	1.11 (360)	&	18.82$\pm$0.27	\\
35014204	&	2015-04-13	&	829.443	&	1.2$^{+0.04}_{-0.05}$	&	0.33$\pm$0.03	&	1.15 (431)		&	1.45$\pm$0.04	&	2.56$\pm$0.03	&	1.16 (431)	&	19.35$\pm$0.22	\\
35014208	&	2015-04-17	&	619.617	&	0.62$^{+0.11}_{-0.15}$	&	0.18$\pm$0.05	&	1.09 (335)		&	1.57$\pm$0.1	&	2.42$\pm$0.03	&	1.09 (335)	&	14.31$\pm$0.22	\\
35014212	&	2015-04-22	&	954.599	&	0.66$^{+0.06}_{-0.08}$	&	0.24$\pm$0.03	&	0.97 (410)		&	1.51$\pm$0.06	&	2.49$\pm$0.02	&	0.98 (410)	&	15.81$\pm$0.17	\\
35014214	&	2015-04-29	&	969.605	&	0.43$^{+0.06}_{-0.08}$	&	0.26$\pm$0.04	&	0.98 (365)		&	1.41$\pm$0.11	&	2.52$\pm$0.02	&	1.01 (365)	&	12.65$\pm$0.15	\\
92204003	&	2015-05-05	&	1020.032	&	0.76$^{+0.08}_{-0.11}$	&	0.21$\pm$0.04	&	1 (380)		&	1.56$\pm$0.07	&	2.44$\pm$0.03	&	1.01 (380)	&	11.82$\pm$0.15	\\
35014216	&	2015-05-06	&	1019.626	&	0.75$^{+0.12}_{-0.17}$	&	0.14$\pm$0.04	&	0.98 (384)		&	1.64$\pm$0.08	&	2.36$\pm$0.04	&	0.99 (384)	&	11.59$\pm$0.16	\\
35014218	&	2015-05-08	&	1094.61	&	0.50$^{+0.08}_{-0.10}$	&	0.28$\pm$0.05	&	1.16 (301)		&	1.4$\pm$0.13	&	2.54$\pm$0.02	&	1.17 (301)	&	6.47$\pm$0.1	\\
35014219	&	2015-05-10	&	1079.617	&	0.44$^{+0.09}_{-0.12}$	&	0.22$\pm$0.05	&	1.12 (318)		&	1.49$\pm$0.13	&	2.47$\pm$0.02	&	1.13 (318)	&	7.49$\pm$0.11	\\
35014226	&	2015-05-24	&	464.195	&	0.7$^{+0.06}_{-0.08}$	&	0.44$\pm$0.07	&	1.11 (245)		&	1.33$\pm$0.11	&	2.68$\pm$0.04	&	1.14 (245)	&	9.27$\pm$0.18	\\
92204006	&	2015-06-02	&	908.953	&	0.7$^{+0.05}_{-0.06}$	&	0.33$\pm$0.04	&	1.03 (355)		&	1.41$\pm$0.07	&	2.58$\pm$0.02	&	1.04 (355)	&	12.43$\pm$0.16	\\
35014242	&	2015-12-14	&	1069.451	&	0.68$^{+0.03}_{-0.04}$	&	0.47$\pm$0.04	&	1.04 (391)		&	1.26$\pm$0.06	&	2.71$\pm$0.02	&	1.07 (391)	&	14.41$\pm$0.14	\\
35014243	&	2015-12-17	&	1014.618	&	0.45$^{+0.06}_{-0.07}$	&	0.27$\pm$0.04	&	1.13 (376)		&	1.41$\pm$0.1	&	2.53$\pm$0.02	&	1.15 (376)	&	12.69$\pm$0.14	\\
34228001	&	2015-12-21	&	2009.225	&	0.63$^{+0.09}_{-0.11}$	&	0.15$\pm$0.03	&	1.1 (466)		&	1.62$\pm$0.06	&	2.38$\pm$0.02	&	1.1 (466)	&	10.64$\pm$0.11	\\
35014251	&	2015-12-27	&	984.454	&	0.68$^{+0.05}_{-0.07}$	&	0.31$\pm$0.04	&	1.16 (371)		&	1.42$\pm$0.07	&	2.56$\pm$0.02	&	1.17 (371)	&	12.17$\pm$0.15	\\
34228006	&	2015-12-30	&	989.62	&	0.98$^{+0.04}_{-0.06}$	&	0.27$\pm$0.03	&	1.18 (464)		&	1.5$\pm$0.04	&	2.51$\pm$0.02	&	1.19 (464)	&	20.61$\pm$0.2	\\
34228007	&	2015-12-31	&	949.447	&	0.73$^{+0.05}_{-0.06}$	&	0.26$\pm$0.03	&	1.19 (443)		&	1.5$\pm$0.05	&	2.5$\pm$0.02	&	1.17 (443)	&	21.09$\pm$0.21	\\
34228009	&	2016-01-03	&	1069.614	&	0.46$^{+0.03}_{-0.04}$	&	0.4$\pm$0.03	&	1.18 (381)		&	1.18$\pm$0.09	&	2.67$\pm$0.01	&	1.2 (381)	&	14.65$\pm$0.14	\\
35014254	&	2016-01-04	&	984.66	&	0.48$^{+0.03}_{-0.05}$	&	0.35$\pm$0.03	&	1.08 (404)		&	1.3$\pm$0.08	&	2.62$\pm$0.01	&	1.15 (404)	&	17.75$\pm$0.17	\\
92399004-Seg1	&	2016-05-26	&	597	&	1.31$^{+0.11}_{-0.12}$	&	0.2$\pm$0.04	&	1.09 (368)		&	1.57$\pm$0.05	&	2.43$\pm$0.05	&	1.09 (368)	&	16.35$\pm$0.24	\\
92399004-Seg2	&	2016-05-26	&	541.6	&	1.29$^{+0.13}_{-0.13}$	&	0.18$\pm$0.04	&	1.19 (354)		&	1.59$\pm$0.06	&	2.41$\pm$0.05	&	1.19 (354)	&	16.15$\pm$0.25	\\
92399006-Seg1	&	2016-06-09	&	495.6	&	1.89$^{+0.28}_{-0.20}$	&	0.21$\pm$0.05	&	1.19 (337)		&	1.56$\pm$0.05	&	2.44$\pm$0.07	&	1.19 (337)	&	28.95$\pm$0.49	\\
92399006-Seg2	&	2016-06-09	&	492.4	&	1.82$^{+0.30}_{-0.20}$	&	0.19$\pm$0.05	&	1.06 (327)		&	1.58$\pm$0.05	&	2.42$\pm$0.07	&	1.06 (327)	&	27.31$\pm$0.49	\\
34228080	&	2016-06-12	&	909.621	&	0.75$^{+0.09}_{-0.12}$	&	0.18$\pm$0.04	&	1.02 (400)		&	1.59$\pm$0.07	&	2.41$\pm$0.03	&	1.03 (400)	&	13.85$\pm$0.17	\\
34228108	&	2017-01-02	&	899.461	&	0.85$^{+0.07}_{-0.09}$	&	0.24$\pm$0.04	&	0.9 (363)		&	1.52$\pm$0.06	&	2.48$\pm$0.03	&	0.9 (363)	&	11.5$\pm$0.16	\\
34228109	&	2017-01-03	&	974.617	&	0.43$^{+0.08}_{-0.08}$	&	0.21$\pm$0.05	&	1 (309)		&	1.46$\pm$0.08	&	2.46$\pm$0.02	&	0.99 (309)	&	7.91$\pm$0.12	\\
34228110-Orb3	&	2017-01-04	&	848.3	&	1.37$^{+0.1}_{-0.09}$	&	0.26$\pm$0.04	&	1.17 (354)		&	1.51$\pm$0.04	&	2.49$\pm$0.05	&	1.18 (354)	&	10.78$\pm$0.16	\\
34228110-Orb5	&	2017-01-04	&	785.3	&	1.25$^{+0.1}_{-0.09}$	&	0.26$\pm$0.05	&	1.01 (344)		&	1.51$\pm$0.05	&	2.49$\pm$0.05	&	1.02 (344)	&	10.79$\pm$0.17	\\
34228116	&	2017-01-10	&	964.617	&	0.79$^{+0.08}_{-0.11}$	&	0.2$\pm$0.04	&	1.2 (390)		&	1.57$\pm$0.07	&	2.43$\pm$0.03	&	1.2 (390)	&	13.38$\pm$0.17	\\
34228118	&	2017-01-19	&	1114.621	&	0.92$^{+0.08}_{-0.10}$	&	0.22$\pm$0.04	&	1.19 (369)		&	1.54$\pm$0.06	&	2.46$\pm$0.03	&	1.2 (369)	&	10$\pm$0.13	\\
81926001	&	2017-01-31	&	1008	&	1.32$^{+0.19}_{-0.17}$	&	0.13$\pm$0.04	&	1.05 (379)		&	1.66$\pm$0.05	&	2.34$\pm$0.05	&	1.05 (379)	&	10.57$\pm$0.15	\\
34228128	&	2017-02-01	&	844.61	&	0.85$^{+0.32}_{-0.64}$	&	0.07$\pm$0.05	&	1.03 (304)		&	1.75$\pm$0.15	&	2.25$\pm$0.07	&	1.03 (304)	&	8.91$\pm$0.17	\\
34228127	&	2017-02-01	&	844.619	&	0.76$^{+0.11}_{-0.12}$	&	0.08$\pm$0.04	&	1.02 (356)		&	1.74$\pm$0.11	&	2.26$\pm$0.05	&	1.02 (356)	&	11.55$\pm$0.18	\\
34228129	&	2017-02-01	&	964.442	&	0.74$^{+0.15}_{-0.16}$	&	0.12$\pm$0.04	&	1.06 (372)		&	1.67$\pm$0.09	&	2.33$\pm$0.04	&	1.06 (372)	&	11.36$\pm$0.16	\\
34228126	&	2017-02-01	&	789.617	&	0.91$^{+0.16}_{-0.16}$	&	0.12$\pm$0.04	&	1.16 (346)		&	1.67$\pm$0.09	&	2.33$\pm$0.05	&	1.16 (346)	&	11.3$\pm$0.18	\\
34228131	&	2017-02-01	&	1049.157	&	0.49$^{+0.10}_{-0.10}$	&	0.18$\pm$0.04	&	0.97 (346)		&	1.56$\pm$0.11	&	2.42$\pm$0.03	&	0.97 (346)	&	9.4$\pm$0.13	\\
34228159	&	2017-03-14	&	1034.621	&	0.43$^{+0.08}_{-0.08}$	&	0.19$\pm$0.04	&	1.01 (336)		&	1.5$\pm$0.11	&	2.44$\pm$0.02	&	1.01 (336)	&	10$\pm$0.14	\\
34228161	&	2017-03-19	&	1004.621	&	0.65$^{+0.11}_{-0.15}$	&	0.15$\pm$0.04	&	1.02 (373)		&	1.62$\pm$0.08	&	2.38$\pm$0.03	&	1.03 (373)	&	12.64$\pm$0.17	\\
31630004	&	2017-04-15	&	864.515	&	1.53$^{+0.08}_{-0.08}$	&	0.3$\pm$0.04	&	1.1 (398)		&	1.47$\pm$0.04	&	2.53$\pm$0.04	&	1.1 (398)	&	16.81$\pm$0.22	\\
31630007	&	2017-04-18	&	904.617	&	0.57$^{+0.09}_{-0.12}$	&	0.21$\pm$0.04	&	1 (322)		&	1.54$\pm$0.1	&	2.45$\pm$0.03	&	1.01 (322)	&	9.35$\pm$0.14	\\
31630009	&	2017-04-21	&	1008.341	&	0.44$^{+0.08}_{-0.08}$	&	0.17$\pm$0.04	&	0.95 (357)		&	1.54$\pm$0.12	&	2.42$\pm$0.02	&	0.95 (357)	&	10.68$\pm$0.14	\\
92412001	&	2017-04-23	&	1173.083	&	0.97$^{+0.08}_{-0.10}$	&	0.22$\pm$0.04	&	1.06 (358)		&	1.55$\pm$0.06	&	2.45$\pm$0.04	&	1.07 (358)	&	10.92$\pm$0.16	\\
31630010	&	2017-04-24	&	889.621	&	0.73$^{+0.11}_{-0.15}$	&	0.18$\pm$0.04	&	1.02 (346)		&	1.59$\pm$0.08	&	2.41$\pm$0.04	&	1.02 (346)	&	10$\pm$0.15	\\
93249001	&	2017-04-25	&	1119.607	&	0.75$^{+0.07}_{-0.10}$	&	0.23$\pm$0.04	&	1.12 (376)		&	1.54$\pm$0.07	&	2.47$\pm$0.03	&	1.13 (376)	&	10.67$\pm$0.14	\\
93249002	&	2017-04-26	&	1009.61	&	0.56$^{+0.14}_{-0.19}$	&	0.15$\pm$0.05	&	1.05 (314)		&	1.61$\pm$0.12	&	2.38$\pm$0.03	&	1.05 (314)	&	10.72$\pm$0.18	\\
93249004	&	2017-04-28	&	1029.447	&	0.49$^{+0.09}_{-0.12}$	&	0.21$\pm$0.04	&	1.08 (334)		&	1.53$\pm$0.11	&	2.45$\pm$0.02	&	1.1 (334)	&	9.03$\pm$0.13	\\
92412002	&	2017-04-30	&	1029.46	&	0.54$^{+0.14}_{-0.18}$	&	0.13$\pm$0.04	&	0.98 (366)		&	1.66$\pm$0.1	&	2.35$\pm$0.03	&	0.98 (366)	&	10.86$\pm$0.15	\\
93249005	&	2017-05-01	&	959.454	&	0.58$^{+0.12}_{-0.16}$	&	0.15$\pm$0.04	&	1.13 (376)		&	1.63$\pm$0.09	&	2.37$\pm$0.03	&	1.14 (376)	&	12.57$\pm$0.17	\\
92412004-Seg1	&	2017-05-14	&	597.6	&	1.23$^{+0.29}_{-0.28}$	&	0.11$\pm$0.05	&	1.14 (320)		&	1.7$\pm$0.08	&	2.3$\pm$0.07	&	1.14 (320)	&	11.13$\pm$0.2	\\
31630014	&	2017-05-22	&	999.607	&	1.04$^{+0.10}_{-0.12}$	&	0.19$\pm$0.04	&	1.08 (361)		&	1.59$\pm$0.06	&	2.42$\pm$0.04	&	1.08 (361)	&	10$\pm$0.14	\\
92412006	&	2017-05-28	&	908.952	&	3.95$^{+0.98}_{-0.92}$	&	0.12$\pm$0.04	&	1.19 (425)		&	1.65$\pm$0.03	&	2.38$\pm$0.1	&	1.18 (425)	&	14.12$\pm$0.19	\\
93249012	&	2017-06-27	&	1084.39	&	0.54$^{+0.09}_{-0.12}$	&	0.2$\pm$0.04	&	1.16 (335)		&	1.56$\pm$0.1	&	2.44$\pm$0.03	&	1.18 (335)	&	8.46$\pm$0.12	\\
31202147	&	2017-12-19	&	1079.11	&	0.6$^{+0.08}_{-0.10}$	&	0.18$\pm$0.03	&	1.15 (434)		&	1.6$\pm$0.07	&	2.41$\pm$0.02	&	1.17 (434)	&	16.32$\pm$0.17	\\
31202212-Seg1	&	2018-03-14	&	495.8	&	1.17$^{+0.18}_{-0.21}$	&	0.13$\pm$0.04	&	1.08 (348)		&	1.67$\pm$0.07	&	2.33$\pm$0.05	&	1.08 (348)	&	19.37$\pm$0.3	\\
31202212-Seg2	&	2018-03-14	&	499	&	1.27$^{+0.16}_{-0.17}$	&	0.15$\pm$0.04	&	1.04 (348)		&	1.64$\pm$0.06	&	2.36$\pm$0.05	&	1.04 (348)	&	20$\pm$0.31	\\

\hline
\end{tabular}
\end{center}
\end{table*}

\begin{table*}
\begin{center}
\label{tab:continued}         
\begin{tabular}{l c c c c c c c c c }
\hline
 ObsID	&	Date of Obs.	&	Exposure	&		&	\emph{eplogpar}	&		&		&	\emph{sbpl}	&	 & 	Flux$_{0.3-10.0\,\rm kev}$\\
 
 \cline{4-6}
 \cline{7-9}
 
  	&		&	(sec)	&	$\epsilon_p$\,(keV)	&	$\beta$	&	$\tilde{\chi_{red}^2}$\,(dof)	&	$\Gamma_{\rm low}$	&	$\Gamma_{\rm high}$		&	$\tilde{\chi_{red}^2}$\,(dof) & ( $10^{-10} \rm erg\,cm^{-2} s^{-1}$)	\\

\hline
\hline
31202213	&	2018-03-15	&	279.457	&	1.07$^{+0.13}_{-0.15}$	&	0.22$\pm$0.06	&	0.97 (271)		&	1.55$\pm$0.09	&	2.45$\pm$0.06	&	0.97 (271)	&	17.32$\pm$0.37	\\
31202214	&	2018-03-16	&	394.713	&	1.05$^{+0.23}_{-0.31}$	&	0.11$\pm$0.05	&	1.09 (306)		&	1.69$\pm$0.1	&	2.31$\pm$0.06	&	1.09 (306)	&	16.81$\pm$0.31	\\
31202215-Seg1	&	2018-03-17	&	596.3	&	0.88$^{+0.07}_{-0.10}$	&	0.29$\pm$0.05	&	1.17 (311)		&	1.48$\pm$0.07	&	2.53$\pm$0.04	&	1.18 (311)	&	12.02$\pm$0.19	\\
31202215-Seg-2	&	2018-03-17	&	454.6	&	0.86$^{+0.09}_{-0.11}$	&	0.29$\pm$0.06	&	1.11 (279)		&	1.48$\pm$0.09	&	2.52$\pm$0.04	&	1.12 (279)	&	12.02$\pm$0.22	\\
31202216	&	2018-03-18	&	1074.608	&	1.13$^{+0.11}_{-0.13}$	&	0.24$\pm$0.06	&	1.04 (296)		&	1.54$\pm$0.07	&	2.47$\pm$0.05	&	1.05 (296)	&	12.78$\pm$0.24	\\
31202231-Seg1	&	2018-04-18	&	495.2	&	1.33$^{+0.28}_{-0.33}$	&	0.12$\pm$0.06	&	1.05 (293)		&	1.68$\pm$0.08	&	2.32$\pm$0.08	&	1.05 (293)	&	10.89$\pm$0.22	\\
31202231-Seg2	&	2018-04-18	&	517.4	&	1.06$^{+0.25}_{-0.20}$	&	0.14$\pm$0.06	&	1.18 (296)		&	1.65$\pm$0.08	&	2.35$\pm$0.06	&	1.18 (296)	&	10.76$\pm$0.21	\\
31202239-Orb1	&	2018-05-20	&	522.1	&	0.62$^{+0.11}_{-0.10}$	&	0.36$\pm$0.08	&	1.12 (224)		&	1.36$\pm$0.14	&	2.61$\pm$0.04	&	1.12 (224)	&	7.34$\pm$0.17	\\
94151014	&	2018-11-28	&	1054.619	&	0.53$^{+0.06}_{-0.08}$	&	0.26$\pm$0.04	&	1.09 (373)		&	1.47$\pm$0.08	&	2.51$\pm$0.02	&	1.11 (373)	&	12.13$\pm$0.14	\\
31630030	&	2019-01-23	&	969.617	&	0.57$^{+0.09}_{-0.10}$	&	0.1$\pm$0.04	&	1.16 (339)		&	1.71$\pm$0.12	&	2.3$\pm$0.04	&	1.17 (339)	&	10$\pm$0.16	\\
31630042	&	2019-02-24	&	944.445	&	0.48$^{+0.07}_{-0.09}$	&	0.21$\pm$0.04	&	1.08 (395)		&	1.54$\pm$0.09	&	2.46$\pm$0.02	&	1.11 (395)	&	15.16$\pm$0.17	\\
31630045	&	2019-03-02	&	1019.617	&	0.78$^{+0.07}_{-0.09}$	&	0.2$\pm$0.03	&	1.13 (428)		&	1.57$\pm$0.06	&	2.43$\pm$0.03	&	1.14 (428)	&	16.73$\pm$0.18	\\
31630046	&	2019-03-04	&	1109.617	&	0.79$^{+0.10}_{-0.13}$	&	0.15$\pm$0.03	&	0.98 (431)		&	1.63$\pm$0.06	&	2.37$\pm$0.03	&	0.98 (431)	&	14.22$\pm$0.16	\\
31630052	&	2019-03-23	&	409.617	&	1.5$^{+0.22}_{-0.18}$	&	0.19$\pm$0.06	&	1.16 (298)		&	1.58$\pm$0.07	&	2.42$\pm$0.07	&	1.16 (298)	&	13.68$\pm$0.27	\\
31630053	&	2019-03-25	&	409.621	&	1.84$^{+0.43}_{-0.25}$	&	0.15$\pm$0.05	&	1.01 (341)		&	1.64$\pm$0.06	&	2.36$\pm$0.07	&	1.01 (341)	&	24.82$\pm$0.43	\\
31630054-Orb1	&	2019-03-28	&	695.2	&	0.74$^{+0.12}_{-0.17}$	&	0.16$\pm$0.04	&	1.02 (338)		&	1.62$\pm$0.09	&	2.39$\pm$0.04	&	1.03 (338)	&	15.57$\pm$0.24	\\
31630054-Orb2	&	2019-03-28	&	233.5	&	1.23$^{+0.16}_{-0.16}$	&	0.22$\pm$0.07	&	1.09 (265)		&	1.56$\pm$0.08	&	2.44$\pm$0.07	&	1.09 (265)	&	22.55$\pm$0.51	\\
31630055	&	2019-03-29	&	883.052	&	1.92$^{+0.17}_{-0.14}$	&	0.26$\pm$0.04	&	1.06 (386)		&	1.5$\pm$0.04	&	2.5$\pm$0.05	&	1.06 (386)	&	18.89$\pm$0.27	\\
31630078	&	2019-06-06	&	939.617	&	0.44$^{+0.09}_{-0.12}$	&	0.21$\pm$0.04	&	1.08 (321)		&	1.52$\pm$0.12	&	2.46$\pm$0.02	&	1.1 (321)	&	9.31$\pm$0.13	\\
11445001	&	2019-06-13	&	999.621	&	0.75$^{+0.09}_{-0.13}$	&	0.17$\pm$0.04	&	1.02 (393)		&	1.6$\pm$0.07	&	2.4$\pm$0.03	&	1.03 (393)	&	12.23$\pm$0.15	\\
11445006	&	2019-06-22	&	1009.871	&	1.16$^{+0.08}_{-0.09}$	&	0.21$\pm$0.04	&	1.04 (406)		&	1.56$\pm$0.05	&	2.44$\pm$0.04	&	1.04 (406)	&	12.74$\pm$0.16	\\
\hline
\end{tabular}
\end{center}
\end{table*}